\documentclass[lettersize, journal]{IEEEtran}
\usepackage{amsmath,amsfonts}
\usepackage{algorithmic}
\usepackage{algorithm}
\usepackage{array}
\usepackage{multirow}
\usepackage{tcolorbox}
\usepackage[caption=false,font=footnotesize]{subfig}
\usepackage{textcomp}
\usepackage{stfloats}
\usepackage{todonotes}
\usepackage{verbatim}
\usepackage{graphicx}
\usepackage{xspace}
\usepackage{cite}
\usepackage{url}
\usepackage[hidelinks]{hyperref}
\usepackage[shortlabels]{enumitem} 
\newcommand{\systemname}{ROAST\xspace}
\begin{document}

\title{\systemname: Risk-Aware Selective Training of Anomaly Detectors Against Evasion Attacks}

\author{Mohammed Elnawawy, Gargi Mitra, Shahrear Iqbal, and Karthik Pattabiraman%
\thanks{M. Elnawawy, G. Mitra, and K. Pattabiraman are with the
Department of Electrical and Computer Engineering, University of
British Columbia, Vancouver, BC, Canada (e-mail: mnawawy@ece.ubc.ca;
gargi@ece.ubc.ca; karthikp@ece.ubc.ca).}%
\thanks{S. Iqbal is with the National Research Council Canada
(e-mail: shahrear.iqbal@nrc-cnrc.gc.ca).}%
\thanks{This project is supported by collaborative research funding from
the National Research Council of Canada's Digital Health and Geospatial
Analytics Program, UBC Four-Year Fellowships (FYF), and the Natural
Sciences and Engineering Research Council of Canada (NSERC). This work has been submitted to IEEE for possible publication. Copyright may be transferred without notice, after which this version may no longer be accessible.}%
\thanks{Corresponding author: Mohammed Elnawawy.}}

% The paper headers
% \markboth{Journal of \LaTeX\ Class Files,~Vol.~14, No.~8, August~2021}%
% {Shell \MakeLowercase{\textit{et al.}}: A Sample Article Using IEEEtran.cls for IEEE Journals}

% \IEEEpubid{0000--0000/00\$00.00~\copyright~2021 IEEE}
% Remember, if you use this you must call \IEEEpubidadjcol in the second
% column for its text to clear the IEEEpubid mark.

\newcommand{\karthik}[1]{\todo[size=\small]{ #1}}
\newcommand{\gargi}[1]{\todo[color=yellow, size=\small]{ #1}}
\newcommand{\shahrear}[1]{\todo[size=\small]{ #1}}
\newcommand{\nawawy}[1]{\todo[color=green, size=\small]{ #1}}
\newcommand{\feedback}[1]{{\color{red} #1}}

\newcommand{\insightbox}[2]{
	\begin{tcolorbox}[
		colframe=blue!40!black, 
		colback=blue!5!white, 
		left=5px,right=5px,top=3px,bottom=3px,
		title=Key Findings: #1]
		#2
	\end{tcolorbox}
}

\maketitle

\begin{abstract}
Safety-critical domains such as healthcare rely on deep neural networks (DNNs) for decision making, yet DNNs remain vulnerable to evasion attacks. Anomaly detectors (ADs) are widely used to protect DNNs, but conventional ADs are trained indiscriminately on benign data from all patients, overlooking cross-patient physiological heterogeneity that can obscure adversarial patterns and reduce recall. In this paper, we propose \systemname, a novel risk-aware outlier exposure (OE) framework that selectively trains ADs to improve recall while largely preserving precision. \systemname identifies patients with lower attack-induced impact scores and focuses AD training on their statistically more regular traces, thereby reducing false negatives. To preserve precision, the framework applies OE by injecting adversarial samples from these patients into the training set while excluding more heterogeneous traces from the other patient cluster. On average, compared with indiscriminate training, \systemname increases recall by 16.2\% in the black-box setting and 3.93\% in the white-box setting, and reduces AD training time by 88.3\% per training cycle, excluding \systemname's one-time attack-simulation, risk-profiling, and clustering overhead, while largely preserving precision.
\end{abstract}


\section{Introduction}
\label{Section: Introduction}
%In recent years, %machine learning (ML) techniques, especially 

Deep neural networks (DNNs) are extensively used in safety-critical applications such as healthcare, 
%~\cite{bhowmik2022deep, alanazi2022using},
%~\cite{alanazi2022using, habehh2021machine, wiens2018machine, bhowmik2022deep, AI-Rad},
% They are used for a wide range of applications, 
including medical imaging and personalized medicine~\cite{bohmrah2025advanced, ahmed2022enhanced}. 
%~\cite{bohmrah2025advanced, panahi2025deep, ji2025continuous, ahmed2022enhanced}. 
However, DNNs are susceptible to 
%~\cite{fawaz2019adversarial},
%~\cite{narodytska2017simple, fawaz2019adversarial, amini2024fast}, 
% especially 
evasion attacks
%~\cite{DBLP:journals/corr/KurakinGB16a},
%~\cite{DBLP:journals/corr/KurakinGB16a, chernikova2019self, herath2021real}, 
%which are known for their ease of execution 
% {\color{red}prevalent since they are relatively easy to execute}\gargi{Paraphrase: well-known for their ease of execution} 
~\cite{goodfellow2015explaining, carlini2017towards}.
% ~\cite{Boesch_2024}.
%~\cite{Boesch_2024, farinetti2018evasion}. 
In evasion attacks, an adversary introduces minor perturbations to the input at inference time so that the DNN 
% misclassifies or 
mispredicts the adversarial sample, 
% {\color{red}a DNN is tricked into misclassifying an adversarial sample at inference time}\gargi{Paraphrase: an adversary introduces minor modifications to input data so that the DNN is forced to misclassify or mispredict}, 
leading to 
% poor accuracy and 
undesirable consequences for safety-critical applications~\cite{goodfellow2015explaining}.
%~\cite{szegedy2014intriguing, goodfellow2015explaining, levy2022personalized}.
In healthcare, evasion attacks may lead to misdiagnoses of patients' future conditions. For example, several DNN-based COVID-19 diagnostic systems were shown to be susceptible to evasion attacks, producing incorrect diagnoses with high confidence~\cite{rahman2020adversarial}. \emph{We focus on protecting healthcare DNN applications against evasion attacks.}
Anomaly detectors (ADs) detect distributional outliers and can
% , when adversarial examples are anomalous, 
serve as a preprocessing defense,
% are techniques 
thereby improving DNN robustness against evasion attacks. Unlike techniques such as adversarial training~\cite{van2022defending}, training-dataset strengthening~\cite{xie2019feature}, and model-algorithm enhancement~\cite{huang2022adversarial}, which often reduce model performance on benign data~\cite{dong2024survey}, ADs act as independent sanity checkers that verify inputs before the DNN processes them, without altering the DNN's functionality~\cite{li2019mad}.
%~\cite{tschuchnig2021anomaly, li2019mad}.
In healthcare, however, this screening layer must balance attack detection against false alarms since false negatives (FNs), or missed attacks, can allow adversarially induced incorrect predictions to reach the clinical workflow, whereas false positives (FPs) can be associated with slower responses to subsequent actionable alerts and denial of service~\cite{rahman2020adversarial,drew2014alarm,bonafide2015nonactionable}. At the same time, healthcare observations are commonly multivariate and temporal. Conventional univariate threshold-based alarms ignore relationships among simultaneously monitored physiological signals and can consequently generate high FP rates~\cite{waterhouse2010multivariate,aboukhalil2008reducing}. However, data-driven ADs can incorporate cross-signal and temporal context, making them suitable for distinguishing anomalous inputs from legitimate physiological variation~\cite{clifton2014predictive}.

Our goal is to build more robust ADs that detect adversarial samples before they reach the DNN, thereby protecting it against evasion attacks.
% General-purpose anomaly ADs
% \gargi{Can we say "general-purpose anomaly ADs?"} 
% are often trained indiscriminately on the benign data of the entire patient dataset to capture the full spectrum of benign patient behavior 
% possible risk scenarios \gargi{Not clear what "risk scenario" refers to. Needs paraphrasing. Are you talking about the full spectrum of benign system behavior?} 
%~\cite{newaz2020adversarial, li2021defending, joe2022machine}. However, human physiology varies significantly among patients.
General-purpose ADs are often trained indiscriminately on large datasets covering a diverse range of scenarios~\cite{joe2022machine}. 
%~\cite{newaz2020adversarial, li2021defending, joe2022machine}.
%While this has been useful in most domains, adopting this approach in 
This approach presents unique problems in healthcare because human physiology varies significantly among patients, producing heterogeneous patient traces. Extreme or irregular observations may reflect genuine pathological states, sensor artifacts, or external and lifestyle factors such as drug intake, diet, physical activity, and stress.
%~\cite{salgado2016noise}.
% \textbf{Define what noisy data is in this context}.
% \gargi{Text in red: Suggested paraphrasing. Please check and incorporate the necessary changes.}
This variation creates three main problems.
\textit{First}, substantial heterogeneity in the training data
% tends to reduce AD robustness, as it 
can obscure meaningful patterns needed for identifying adversarial data, ultimately causing a decline in the AD's recall~\cite{GUPTA2019466}.
%~\cite{GUPTA2019466, shen2022adversarial}.
% samples from patients who are highly vulnerable \gargi{GM: This looks abrupt - Who is a vulnerable patient? Why is their data noisy?}
% \gargi{GM: We should mention recall degradation as a consequence here. Else, it looks abrupt later } 
% \textit{Second}, it degrades the model's generalizability, which is crucial for deploying models in diverse adversarial settings~\cite{alawad2019adversarial}. 
\textit{Second}, processing the full dataset increases training cost, motivating data-selection methods that reduce the training set while seeking to preserve model utility~\cite{benarba2026quantitative}. \textit{Third}, ADs are traditionally trained using in-distribution data that represent benign behavior, which may not generalize to unseen outliers~\cite{hendrycks2019deep}.

We propose \systemname\footnote{
    % We have prepared an anonymized replication package: \url{https://bit.ly/4ck0nzx}. We will make \systemname publicly available if the paper is accepted.
    Code available at: \url{https://github.com/DependableSystemsLab/ROAST.git}
}, a Risk-aware Outlier exposure framework for Adversarial Selective Training of ADs, to address those challenges.
``Risk'' refers to adversarial evasion risk, i.e., a data-driven proxy for how strongly an attack can influence the victim model.
% , rather than a clinical measure of patient harm severity.
% without compromising  their precision. 
We hypothesize that training ADs 
% such as \textit{k}-nearest neighbors (\textit{k}NN), One-Class Support Vector Machines (One-Class SVM), and Multivariate Anomaly Detection for Time-series Data with Generative Adversarial Networks (MAD-GAN)~\cite{li2019mad} 
% \karthik{Why are we naming the ADs here? If so, we should justify why we choose these three ADs. Does this mean our technique won't work with any other AD?}
using benign and adversarial data from patients who are less vulnerable to attacks can enhance their recall 
% lowering the false negative rate 
while reducing the training-set size and computational cost. We define vulnerability as adversarial susceptibility rather than medical instability. Under a fixed, physiologically constrained attack, \systemname constructs a temporal risk profile per patient by tracking empirically weighted attack impact over time. It clusters patients' profiles and labels the cluster with lower aggregate impact as less vulnerable and the other as more vulnerable.

%	In other words, a patient is considered more vulnerable if an adversary can introduce small, semantically valid modifications to their physiological signals while successfully altering the DNN's output as opposed to less vulnerable patients who require larger or less plausible perturbations to induce misclassification.

To power \systemname, we design a risk profiling stage that groups patients by their vulnerability to attacks and selects the less vulnerable ones for training
%  This stage is equivalent to adversarial influence profiling: 
%  It ranks patients by attack susceptibility 
using a data-driven proxy score. % rather than by clinically validated harm severity. 
The intuition is that, within the patient population, data from less vulnerable patients may provide a more regular representation of the benign distribution~\cite{akhtom2024enhancing}. We hypothesize that patients whose risk profiles show lower attack-induced impact also contribute less attack-sensitive data for AD training.
% The evaluation tests this hypothesis rather than assuming that statistical regularity follows from vulnerability by definition.
%~\cite{bosman2025robustness}.
%~\cite{garaev2024not, rodriguez2022role, bosman2025robustness}.

\emph{By training on benign data from the lower-impact patient cluster, the AD excludes some cross-patient heterogeneity that may obscure adversarial patterns, thereby improving recall in our evaluation.}
However, relying solely on their benign data may induce the 
model to learn a very narrow view of what is benign, potentially harming precision~\cite{hendrycks2019deep}. To address this problem, we introduce adversarial examples in a controlled manner to complement the benign data from less vulnerable patients and help the AD learn a richer representation of the sample space through unsupervised learning. The main challenge is to inject these examples without reducing precision. We therefore restrict injection to adversarial examples generated from less vulnerable patients; these examples are structured, near-boundary anomalies rather than arbitrary noise. This approach expands the learned benign region enough to prevent the over-tightening caused by benign-only selective training without producing the overly broad region learned through indiscriminate training.
We evaluate \systemname on three publicly available healthcare time-series datasets: OhioT1DM for blood glucose prediction~\cite{marling2020ohiot1dm}, MIMIC for mortality prediction~\cite{johnson2020mimic},
%~\cite{johnson2020mimic, physionet_mimic},
and PhysioNet CinC for sepsis prediction~\cite{reyna2020early}.
These datasets contain observations from different patients in real-world settings.
%\karthik{Added}
%~\cite{reyna2020early, physionet_sepsis, physionet_mimic}.
% \gargi{We should ideally have one citation for each dataset}. 
Our evaluation spans four adversarial techniques: the Universal Robustness Evaluation Toolkit (URET)~\cite{eykholturet} for black-box attacks and three gradient-based white-box attacks---the Fast Gradient Sign Method (FGSM)~\cite{goodfellow2015explaining}, Projected Gradient Descent (PGD)~\cite{mkadry2017towards}, and Carlini--Wagner (C\&W)~\cite{carlini2017towards}. We evaluate five ADs with complementary detection mechanisms: (1) \textit{k}-nearest neighbors (\textit{k}NN), (2) One-Class Support Vector Machine (One-Class SVM), (3) Multivariate Anomaly Detection with Generative Adversarial Networks (MAD-GAN)~\cite{li2019mad}, (4) Deep Support Vector Data Description (Deep SVDD)~\cite{ruff2018deep}, and (5) an LSTM autoencoder (LSTM-AE)~\cite{malhotra2016lstm}. URET, FGSM, \textit{k}NN, One-Class SVM, and MAD-GAN provide the primary recall and precision results, while the complete suite supports the fairness and cross-attack generalization analyses.

In summary, we make three contributions in this paper. 
% \karthik{Let's use  verbs in the contributions - Develop, Propose etc. Also, not sure why 3 and 4 are contributions. They're more based on the results of our experiments, correct?}
% Point 4 is a contribution, since it does not rely on the results of our experiments. We had to adapt URET for time-series data before it could be applied in our evaluation. I also attempted to emphasize the significance of contribution 3. Perhaps I have not expressed it clearly, but what I mean is that it constitutes a contribution because we are not only proposing the framework itself, but also systematically outlining the prerequisite requirements for the inputs of the framework to guide its users.
\begin{enumerate}
    \item We introduce \systemname, a risk-aware framework that constructs temporal, patient-level risk profiles and partitions patients based on their vulnerability to evasion attacks.
    \item We develop a strategy for selectively training ADs on benign and adversarial data from less vulnerable patients.
    %  identified by \systemname.}
    \item We evaluate \systemname on three real-world healthcare datasets, four attacks, and five AD techniques, and report how its benefits vary across the evaluated detector families and dataset settings.
%    \item Present a motivating case study that underscores the importance of risk profiling in healthcare and characterizing datasets using their statistical properties.
%    \karthik{I don't think this should be claimed as a contribution. Instead, claim the \systemname tool as a contribution and its evaluation.}
    % \item We provide insights into the characteristics of ADs and datasets that maximize \systemname's effectiveness.
    % \item Adapt URET~\cite{eykholturet} to support evasion attacks on time-series data, as healthcare datasets are predominantly time-series in nature. 
    % \karthik{I think we should reframe this in terms of the experiments we did rather than a modification of URET.}
\end{enumerate}

% The results of our experiments show that compared to indiscriminate training, selective training guided by our risk profiling framework achieves a recall increase of up to 27.5\% and 89.8\% on anomaly ADs like \textit{k}NN and One-Class SVM, respectively, with little to no impact on precision. Furthermore, when trained on the less vulnerable patients, a MAD-GAN AD maintains a recall that is as high as one with no change to its precision, and a reduction in training set size of up to 93.2\%. 
% % as opposed to indiscriminately training it on the entire dataset. 
% Therefore, our risk profiling framework helps anomaly ADs achieve higher recalls with minimal impact on precision, while significantly reducing the training set size.
% \karthik{You need to put these numbers in context. What does it mean from a practical perspective in terms of attacks and time? Also, what're the absolute numbers of the recall and precision?}

Our experiments show that, compared to indiscriminate training, \systemname improves AD 
% robustness against adversarial perturbations. The results show a 
recall by 16.2\% (black-box) and 3.93\% (white-box) on average 
%up to 89.8\% (i.e., a recall increase from $\approx$0.499 to $\approx$0.946 on the MIMIC dataset using One-Class SVM), 
while preserving precision for boundary-based and temporal generative detectors; the observed precision loss is confined only to the local, instance-based \textit{k}NN detector.
% , which is only 4.97\%). 
%\karthik{Present the average value not best case.}
% For instance, using the \textit{k}NN AD on the OhioT1DM dataset improves recall from 0.759 to 0.968 (a relative increase of 27.5\%), while One-Class SVM improves recall on the MIMIC dataset from 0.501 to 0.951 (a relative increase of 89.8\%). These gains are achieved with little to no degradation in precision (i.e., in the worst-case scenario, a precision drop from 0.594 to 0.565 - only 4.97\%).
\systemname also requires 75.0\%, 93.2\%, and 72.0\% fewer training samples on OhioT1DM, MIMIC, and PhysioNet CinC, respectively. 
% when trained only on less vulnerable patients, a MAD-GAN AD maintains the same recall levels as training on the entire dataset with no drop in precision, while requiring 75.0\%, 93.2\%, and 72.0\% fewer training samples on the OhioT1DM, MIMIC, and PhysioNet CinC datasets, respectively. 
This reduction translates into AD training-time reductions ranging from 70.66\% to 99.77\% per training cycle, excluding \systemname's one-time attack-simulation, risk-profiling, and clustering overhead.
% for connected healthcare systems. 
% \karthik{Can we quantify this reduction?}
%Therefore, \systemname not only enables ADs to detect a larger number of adversarial inputs but also reduces their training overhead.}
% \gargi{You will give this summary in the results section anyway. So just keep the best results for the Intro.}}

\section{Background}
\label{Section: Background}

\subsection{Evasion Attacks}
\label{Subsection: Evasion_Attacks}
Evasion attacks are inference-time attacks in which an adversary minimally perturbs inputs to induce incorrect DNN decisions~\cite{goodfellow2015explaining}. They can be gradient-based, such as FGSM, PGD, and C\&W~\cite{goodfellow2015explaining, mkadry2017towards, carlini2017towards}, or non-gradient-based, such as reinforcement learning and GAN-based approaches~\cite{sutton2018reinforcement, goodfellow2014generative}.
% , and may operate in black-box, white-box, or gray-box settings~\cite{eykholturet, goodfellow2015explaining, mkadry2017towards, liu2022surrogate}. 
In this work, we evaluate \systemname under two threat models, black-box and white-box. 
For the \emph{black-box} setting, we instantiate attacks using URET, a non-gradient-based, general-purpose evasion-attack framework that manipulates data points independently of the input type or task domain~\cite{eykholturet}. Given a data point and a set of predefined transformations, we use URET's search algorithm to create an adversarial sample that is semantically and functionally valid. For the \emph{white-box} setting, we use (a) FGSM~\cite{goodfellow2015explaining}, a gradient-based attack that perturbs inputs in the direction of the loss gradient,
%  to maximize the model's prediction error
(b) PGD~\cite{mkadry2017towards}, which iteratively applies FGSM-style steps and projects each result back into a bounded perturbation region, and (c) C\&W~\cite{carlini2017towards}, which optimizes a perturbation penalty and an attack objective jointly. 
% We use PGD and C\&W in our cross-attack generalization analysis (Section~\ref{Section: Evaluation}) to test whether \systemname's patient selection transfers to attacks not used during risk profiling.
 %Evaluating against both black-box (URET) and white-box (FGSM) attacks ensures our framework is robust across complementary threat models.

% \karthik{This seems weird. URET is a framework, while FGSM is a technique - they don't parallel each other. What about gray box attacks?}
% \nawawy{We only evaluate against black-box and white-box as they represent the two ends of the spectrum which represent either no knowledge or full model knowledge. I tried incoprporating this into the paragraph by adding the "two endpont threat model" statements. If you think further clarifications is required we could remove gray-box altogether as we never mention them again.}

% \gargi{End it with what type of evasion attack you consider in this work. You can also mention 1/2 sentences about URET.}

\subsection{Datasets}
% \karthik{Maybe this is better expressed in a table?}
\label{Subsection: Datasets}
Table~\ref{Table: Dataset_Summary} summarizes the three datasets used. OhioT1DM has two patient subsets: six patients from 2018 and six from 2020, denoted by \textit{Subset A} and \textit{Subset B}, respectively. We use patient identifiers in patient-level plots and analyses (e.g., A\_3 denotes patient 3 from Subset A). All datasets are publicly available and de-identified. We use them for simulation-based experiments and collect no new human-subject data.

\begin{table}[t]
\centering
\caption{Summary of datasets used in this study.}
\label{Table: Dataset_Summary}
\begin{tabular}{|p{0.16\columnwidth}|p{0.74\columnwidth}|}
\hline
\textbf{Dataset} & \textbf{Summary} \\ \hline
OhioT1DM

\cite{marling2020ohiot1dm} & 
\textit{\textbf{Task}}: blood glucose prediction. 

\textit{\textbf{Cohort}}: 12 type 1 diabetes patients. 

\textit{\textbf{Features}}: CGM glucose, finger sticks, insulin, carbs, etc.
% carbohydrate intake, sleep, activity, etc.

\textit{\textbf{Sampling}}: approximately every five minutes.

\textit{\textbf{Characteristics}}: dense, low-to-moderate-dimensional, and temporally smooth. \\ \hline

MIMIC

\cite{johnson2020mimic} & 
\textit{\textbf{Task}}: mortality prediction. 

\textit{\textbf{Cohort}}: $>$ 65K ICU and 200K emergency patients. 

\textit{\textbf{Features}}: vitals, labs, medications, etc.

\textit{\textbf{Sampling}}: irregular (frequent vitals and intermittent measurements of the remaining variables). 

\textit{\textbf{Characteristics}}: high-dimensional and sparse. \\ \hline

PhysioNet CinC

\cite{reyna2020early} & 
\textit{\textbf{Task}}: sepsis prediction. 

\textit{\textbf{Cohort}}: $>$ 60K ICU patients. 

\textit{\textbf{Features}}: vitals, labs, length of stay, etc.

\textit{\textbf{Sampling}}: hourly recordings with up to 40 variables.

\textit{\textbf{Characteristics}}: high-dimensional and sparse. \\ \hline

\end{tabular}
\end{table}

\subsection{Defenses}
\label{Subsection: Defenses}
ADs can serve as preprocessing defenses against evasion attacks by flagging inputs that deviate from the learned benign distribution~\cite{grosse2017statistical}. 
Prior work has instantiated this idea using local intrinsic dimensionality on DNN activations~\cite{ma2018characterizing}, class-conditional Mahalanobis distance metrics~\cite{lee2018simple}, and input feature squeezing~\cite{xu2018feature}. These methods inspect DNN activations or transformed inputs, whereas \systemname changes the composition of the AD training set. Therefore, these methods are complementary defenses rather than direct training-set comparators.
In this work, for the primary recall and precision evaluation, we evaluate the following baseline ADs:
%  for protecting DNN models against evasion attacks: 
\textit{k}NN, One-Class SVM, and MAD-GAN~\cite{li2019mad}. \textit{k}NN supports local density estimation, One-Class SVM captures rare nonlinear patterns in high-dimensional spaces, and MAD-GAN models multivariate feature dependencies that are useful in safety-critical time-series settings~\cite{li2019mad}. Prior work has used \textit{k}NN and One-Class SVM for anomaly detection in healthcare~\cite{eze2023anomaly}; MAD-GAN is also a suitable baseline because it captures multivariate temporal dependencies common in healthcare data. As a separate robustness extension, we add two deep ADs: Deep SVDD~\cite{ruff2018deep}, a neural one-class boundary that maps inputs into a minimum-volume hypersphere, and an LSTM autoencoder (LSTM-AE)~\cite{malhotra2016lstm}, which scores anomalies using the reconstruction error of an LSTM encoder--decoder.

\subsection{Connected Healthcare Architecture}
\label{Subsection: Connected_Healthcare_Architecture}
\begin{figure}[t]
	\centering
	\includegraphics[scale=0.22]{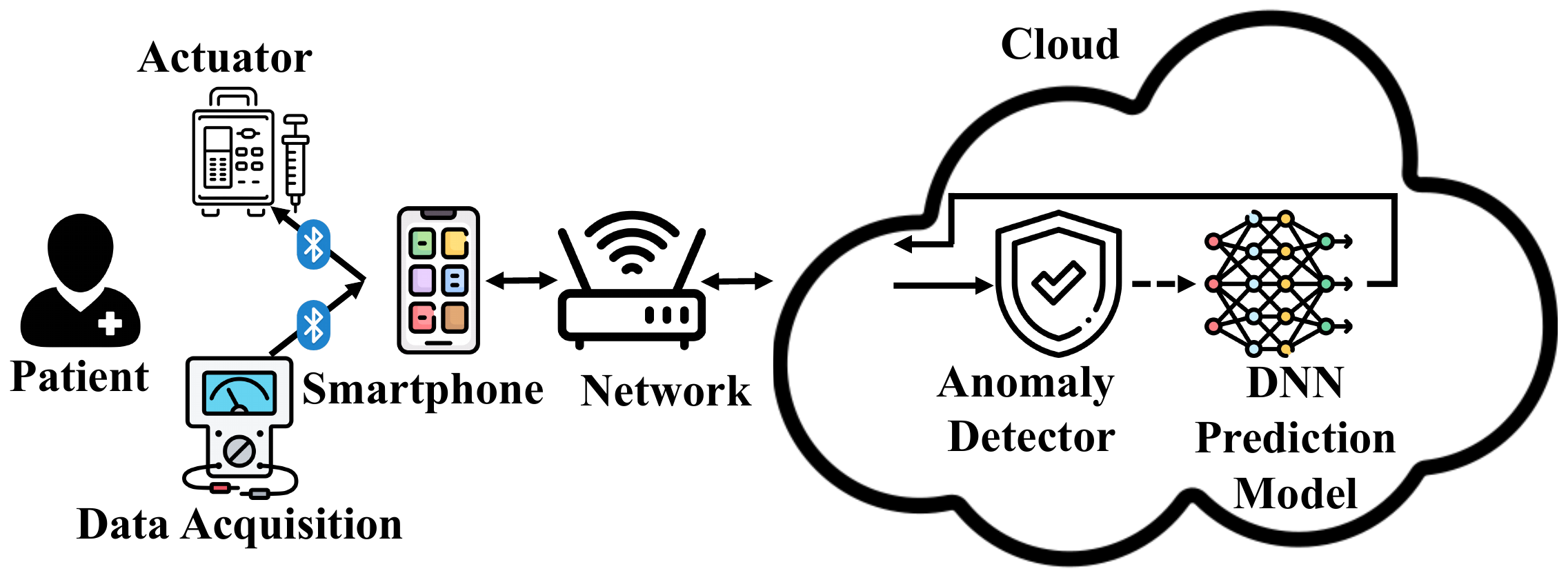}
	\caption{System architecture of the patient monitoring system, comprising a data acquisition device, smartphone, actuator, DNN, and anomaly detector.}
	
	\label{Figure: Architecture}
\end{figure}

% We adopt and extend the system architecture presented in Elnawawy et al.~\cite{elnawawy2024systematically} to simulate evasion attacks on medical systems.\gargi{Do we need to say this? Also, add a sentence on why we consider such an architecture - are they highly susceptible to evasion attacks due to the diversity in deployment environment and high number of peripherals providing input to the ML model?} 
We consider the patient-monitoring system in Figure \ref{Figure: Architecture}. The system comprises a data-acquisition device, such as a continuous glucose monitor (CGM), a heart-rate monitor, or another sensor that collects patients' vitals at regular intervals and transmits them to a mobile app.
% via Bluetooth. 
The app sends the measurements to the cloud, where an AD inspects them for malicious patterns. If a measurement is benign, the system passes it to the DNN for processing; otherwise, the patient may review the flagged measurement, and the data-acquisition device may repeat the measurement. A physiologically plausible adversarial perturbation may appear reasonable to the patient, especially under time pressure or without sufficient clinical context. In the absence of an AD, an adversarial perturbation may reach the DNN and cause a misprediction that results in an incorrect clinical action, such as an erroneous insulin dose or pacemaker rate. The DNN predicts future values, such as glucose or heart rate, and sends them to the app. The app then infers a recommended action, such as an insulin dose or pacemaker-rate adjustment, which the patient can approve before the actuator executes it.

% \karthik{We need to say why the patient can't just check the validity of the action? Why do we need ADs?}

Healthcare DNNs can be categorized as global or patient-specific models. We focus on global models because they capture shared patterns across patients and are therefore more affected by cross-patient variability and evasion attacks than personalized models~\cite{potosnak2025global}.
% \karthik{Need citation}

\subsection{Definition of Risk}
\label{Subsection: Definition_of_Risk}
%Existing work~\cite{allodi2017security} defines risk as the potential for loss or harm when a threat exploits a system vulnerability, and is typically modelled as a function of likelihood and impact.
Following prior work that defines risk as the potential for loss or harm when a threat exploits a system vulnerability~\cite{allodi2017security}, we define risk as the impact of an adversarial input perturbation that shifts a system from a safe state to an unsafe state, which may harm the patient. For instance, if an attacker alters a patient's glucose reading so that the blood glucose management system (BGMS) recommends a high insulin dose, causing a transition from a \textit{safe} state (normal glucose) to an \textit{unsafe} state (hypoglycemia), we consider the attack high risk because of its potential impact on the patient's health.

\subsection{Dataset Distillation and Data Filtering}
\label{Subsection: Dataset_Distillation}
Training on large, heterogeneous datasets is expensive. Dataset distillation compresses training data into smaller, information-rich subsets~\cite{wang2018dataset}. Related approaches use gradient matching~\cite{zhao2021dataset}, factorization~\cite{liu2022dataset}, cleaning~\cite{chandola2009anomaly}, and noise filtering~\cite{chan2025remlx}. However, generic filtering alone is insufficient for robust ADs because ADs are typically trained only on benign data, which can induce a narrow notion of benignity. Researchers have therefore proposed alternative training techniques, including outlier exposure, which exposes ADs to benign and adversarial samples during training to help them learn a richer representation of benign and anomalous samples~\cite{hendrycks2019deep}. \systemname differs from these methods by clustering patients according to attack-induced impact rather than labeling statistically or clinically unusual measurements as noise, making it a robustness-oriented selective-training strategy rather than a generic data-cleaning technique.

%\karthik{I don't understand this last sentence. What's specifically designed?}
% To the best of our knowledge, our selective training technique is the first to adopt outlier exposure using a risk-centric approach that groups training samples into high- and low-risk based on their vulnerability to the evasion attack and employs them to enhance ADs' performance.
% \karthik{First of all, this should go under related work. Also, you need to say why existing ADs are not sufficient.}

\section{Threat Model}
\label{Section: Threat Model}
The attacker's \textit{goal} is to deceive the system so that it produces misleading predictions, alerts, or treatment recommendations. For example, in the case of a BGMS, the system may suggest an unnecessary insulin dose even when direct actuation is protected by safety interlocks. % In the case of mortality or sepsis prediction, the system may suggest a wrong pacemaker rate, leading to under-pacing or over-pacing and potentially heart failure. 
The attacker's \textit{strategy} is to manipulate vitals, such as blood glucose and heart rate, so that they remain within physiologically feasible ranges yet cause erroneous DNN predictions and downstream actions while evading detection.
	% \gargi{I think we are already using the term "normal" in a different context for the BGMS setup. Also, the manipulated values should be within normal physiological ranges. So can we use an alternative word for "abnormal"?}
	% In a BGMS, the adversary manipulates the patient's blood glucose levels to values that exceed 125 mg/dL (hyperglycemic in a fasting state) or 180 mg/dL (hyperglycemic two hours postprandial). In a mortality prediction system, the adversary manipulates the patient's pacemaker rate to values between 150 and 200 bpm (abnormally high). In a sepsis prediction system, the adversary manipulates heart rate to values between 120 and 140 bpm (abnormally high). 
	
	In our threat model, the adversary's \textit{capabilities} enable them to manipulate one or more vital signs. The attacker is assumed to gain write access to the sensor-to-app data path through a feasible cyber entry point in the data-acquisition pipeline, such as a compromised wireless protocol, a Bluetooth exploit, insecure middleware, or malware on an intermediate component~\cite{niu2025securing, rasmussen2022blurtooth}. This abstraction is consistent with reports that connected healthcare sensing devices may expose unencrypted or weakly encrypted communication paths~\cite{niu2025securing}. Moreover, the adversary is assumed to be unaware of the downstream AD and optimizes only the victim-model objective. The adversary does not adapt its perturbations to minimize the AD's anomaly score.
	
	In our experiments, we instantiate this threat model by manipulating CGM measurements, pacemaker rate, and heart rate in OhioT1DM, MIMIC, and PhysioNet CinC, respectively; however, the framework is not limited to these specific channels or to single-feature perturbations.
	% {\color{red}Manipulation of other features remains beyond the attacker's capabilities.}\gargi{Repetitive. Not required.} 
	We also assume that the adversary can compromise the smartphone~\cite{suarez2015compartmentation} and read other features, including those manually entered by the patient, to ensure the physiological plausibility of adversarial samples. For example, an adversary can craft an adversarial heart-rate value while preserving its correlation with features such as blood pressure to reduce the likelihood of detection.
%However, \systemname is designed to be robust even if the threat model considers attacking multiple features.

\section{Motivational Case Study}
\label{Section: Motivational_Case_Study}

The underlying assumption of ADs as evasion detectors is that adversarial examples are anomalous with respect to the benign data distribution.
% {\color{red}usually come from a different distribution than}\gargi{suggested paraphrasing: are anomalous with respect to benign data distribution.} benign data 
%~\cite{grosse2017statistical}. 
Hence, popular ADs
% based on techniques such as \textit{k}NN, One-Class SVM, and MAD-GAN 
work by treating new samples that lie far from the learned benign data as adversarial.
% {\color{green}identifying if a new sample lies far from the learned benign data distribution}\gargi{Paraphrase: treating new samples that lie far from the learned benign data as adversarial}
% {\color{red}, treating adversarial samples as outliers}\gargi{It's actually the reverse, right? They treat outliers as adversarial samples. Anyway this part if redundant IMO}. 
This motivates us to examine whether heterogeneity in benign patient traces is associated with different responses to evasion attacks. 
% Statistical deviation alone, however, does not establish attack vulnerability or clinical risk.
% \gargi{Doubt: If outliers can be easily manipulated to fall \textit{outside} benign decision boundary, they would be easily caught by the ADs, right? IMO, evasion attack is successful when after manipulation, the data point still stays within the decision bdy, but the outcome of ML model is changed. Am I missing something?}.
% \gargi{Not clear how this statement can be derived from the previous ones. Please clarify.} 
To characterize cross-patient physiological heterogeneity, we conduct a preliminary statistical analysis of the percentage of outliers in patients' benign traces. Here, an \emph{outlier} is an observation flagged by a statistical criterion; it does not necessarily indicate measurement error, sensor noise, or an invalid physiological state. Clinically abnormal observations may be genuine and medically important. We use two common outlier-identification techniques that are well suited to non-normally distributed data, which are common in medical datasets~\cite{bono2017non}:
%~\cite{altman1995statistics, choi2022log, bono2017non}:
% \karthik{Need citation}: 
(\textit{a}) the \textit{modified Z-score} ($M_i$) in Equation \ref{Equation: Z-score}, where $x_i$ is the $i^{th}$ observation, $\tilde{x}$ is the dataset median, and $\mathrm{MAD}$ is the median absolute deviation defined in Equation \ref{Equation: MAD}; and (\textit{b}) the \textit{interquartile range} ($IQR$) in Equation \ref{Equation: IQR}, where $Q1$ and $Q3$ are the first and third quartiles, respectively. We classify a data point as an outlier when its absolute modified Z-score exceeds the standard cutoff of 3.5 or when it lies outside the interval [$Q1-1.5 \cdot IQR$, $Q3+1.5 \cdot IQR$].
% \gargi{Need citations for the techniques. Are these popular measures?}
%Yes, these are common techniques and we used common thresholds for our experiments. I don't think a citation is needed.
\begin{equation}
\label{Equation: Z-score}
    M_i = \frac{0.6745 \, (x_i - \tilde{x})}{\mathrm{MAD}}
\end{equation}

\begin{equation}
\label{Equation: MAD}
    \mathrm{MAD} = \mathrm{median}\left( \, \big| x_j - \tilde{x} \big| : j = 1, 2, \dots, n \, \right)
\end{equation}

\begin{equation}
\label{Equation: IQR}
    IQR = Q3 - Q1
\end{equation}

Our motivational case study uses three medical time-series datasets---OhioT1DM, MIMIC, and PhysioNet CinC---to highlight the need for risk profiling in healthcare DNNs. We analyze the proportion of outliers per patient across all features using the modified Z-score and IQR methods to characterize physiological variation among patients. 
% We report only the five highest-ranked patients by outlier proportion
% \gargi{top in terms of what? Importance? Please clarify} 
% features and 
% (in descending order) per dataset in Table
% \ref{Table: Outliers_Per_Feature} and 
For each dataset, Table \ref{Table: Outliers_Per_Patient} reports only the five patients with the highest outlier proportions, in descending order. We also report the average and standard deviation across all patients. We observe 
% that the IQR method flags substantially more outliers than the Z-score method across all three datasets, due to the non-normally distributed nature of the medical datasets. We also notice  
that outliers are distributed non-uniformly because, in most cases, the highest-ranked patients' outlier proportions substantially exceed the corresponding averages across all patients under both methods.
% \gargi{Patient and feature \#5 in OhioT1DM are not far from the average it seems. Should we say `in most cases'?}
% \gargi{Which ones are the overall average numbers?}. 
This result establishes that benign-trace heterogeneity is not uniform across patients, but it does not by itself show that patients with more statistical outliers are easier to attack. It instead motivates the hypothesis that attack-induced impact and benign-trace heterogeneity may be associated. Consequently, training ADs indiscriminately on benign data from all patients may broaden the learned benign region and reduce its ability to distinguish adversarial inputs.
% \karthik{This is wierdly phrased; what do you menan by "not ideal"? What'd you lose as a result. }
Hence, identifying the less (and more) vulnerable patients 
% using risk profiling 
allows us to test whether patient-level selection can reduce training-distribution heterogeneity and improve AD robustness.
%  against evasion attacks.
% \karthik{In general, we should avoid words like "better". Better in what sense? Also, what does overcome mean?}
% \gargi{What do you mean by one-size-fits-all AD? Aren't we still designing a one-size-fits-all AD (i.e., one that gives good results for both low and high-risk patients)? What do you exactly mean by a `better' AD?}.}

%%%%%%%%%%%%%%%%%%%%%%%%%%%%%%%%%%%%%%%%%%%%%%%%%%%%%%%%%%%%%%%%%%%%%%%%%%%%%%%%%%%%%%%%%%%%%%%%%%%%%%%%%%%%%%%%%%%%%%%%%%%
% Please add the following required packages to your document preamble:
% \usepackage{multirow}
% \usepackage{graphicx}
\begin{table}[t]
	\centering
	\caption{Outlier proportions for the five highest-ranked patients in each dataset, ordered from highest to lowest.}
	%  Rows 1--5 denote ranks, not patient identifiers.}
	\label{Table: Outliers_Per_Patient}
	\resizebox{\columnwidth}{!}{%
		\begin{tabular}{|l|ll|ll|ll|}
			\hline
			\multicolumn{1}{|c|}{\multirow{2}{*}{\textbf{Rank}}} &
			\multicolumn{2}{c|}{\textbf{OhioT1DM}} &
			\multicolumn{2}{c|}{\textbf{MIMIC}} &
			\multicolumn{2}{c|}{\textbf{PhysioNet CinC}} \\ \cline{2-7} 
			\multicolumn{1}{|c|}{} &
			\multicolumn{1}{c|}{\textbf{Z-score}} &
			\multicolumn{1}{c|}{\textbf{IQR}} &
			\multicolumn{1}{c|}{\textbf{Z-score}} &
			\multicolumn{1}{c|}{\textbf{IQR}} &
			\multicolumn{1}{c|}{\textbf{Z-score}} &
			\multicolumn{1}{c|}{\textbf{IQR}} \\ \hline
			1                  & \multicolumn{1}{l|}{19.32} & 7.76 & \multicolumn{1}{l|}{23.01} & 20.41 & \multicolumn{1}{l|}{37.20} & 34.96 \\ \hline
			2                  & \multicolumn{1}{l|}{16.93} & 6.89 & \multicolumn{1}{l|}{23.01} & 20.41 & \multicolumn{1}{l|}{34.96} & 34.76 \\ \hline
			3                  & \multicolumn{1}{l|}{15.0}  & 5.28 & \multicolumn{1}{l|}{23.01} & 20.41 & \multicolumn{1}{l|}{30.16} & 30.49 \\ \hline
			4                  & \multicolumn{1}{l|}{14.95} & 3.52 & \multicolumn{1}{l|}{23.01} & 20.41 & \multicolumn{1}{l|}{30.03} & 30.4  \\ \hline
			5                  & \multicolumn{1}{l|}{14.61} & 2.47 & \multicolumn{1}{l|}{23.01} & 20.41 & \multicolumn{1}{l|}{29.72} & 29.02 \\ \hline
			\textbf{Average}   & \multicolumn{1}{l|}{11.94} & 3.22 & \multicolumn{1}{l|}{6.32}  & 3.82  & \multicolumn{1}{l|}{4.77}  & 2.43  \\ \hline
			\textbf{Std. Dev.} & \multicolumn{1}{l|}{4.38}  & 2.23 & \multicolumn{1}{l|}{3.35}  & 3.00  & \multicolumn{1}{l|}{3.74}  & 2.90  \\ \hline
		\end{tabular}%
	}
\end{table}
%%%%%%%%%%%%%%%%%%%%%%%%%%%%%%%%%%%%%%%%%%%%%%%%%%%%%%%%%%%%%%%%%%%%%%%%%%%%%%%%%%%%%%%%%%%%%%%%%%%%%%%%%%%%%%%%%%%%%%%%%%%

\begin{figure}[t]
\centering
\includegraphics[width=0.44\textwidth]{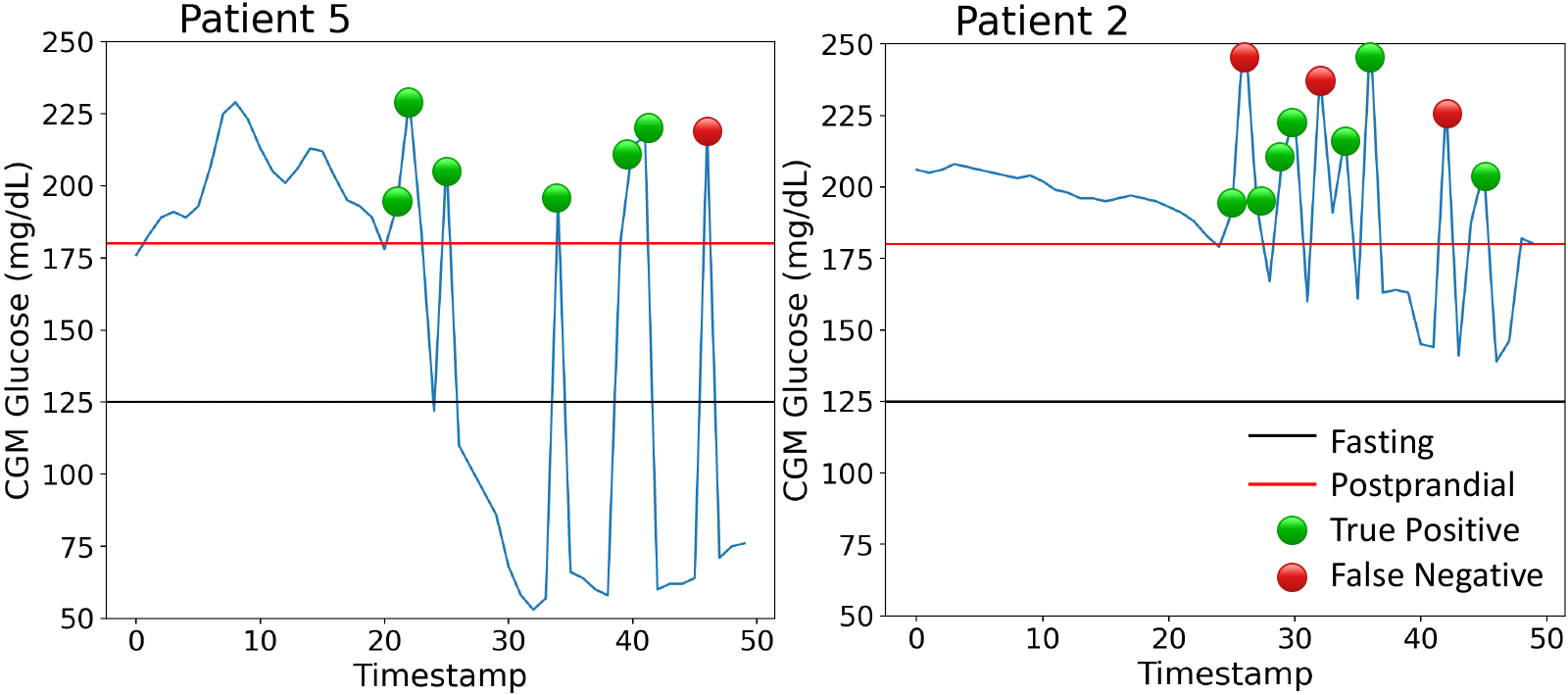}
\caption[Sample Glucose Traces with False Negatives and True Positives]{\textit{k}NN anomaly detection on glucose traces from A\_5 and A\_2 (OhioT1DM), showing a higher FN rate for A\_2 under indiscriminate training. Adversarial samples are generated using URET.}

\label{Figure: Plot_FN_TP_kNN}
\end{figure}

% To further illustrate that indiscriminately training ADs on all patients without considering their vulnerability levels to evasion attacks results in a less robust AD, 
To demonstrate the loss of robustness in ADs,
% that a one-size-fits all solution\gargi{In the intro, we talked about noise removal and outliers. Here, the problem description shifts to one-size-fits-all. Not sure how they are related. We need to maintain consistency across sections and use the same terminology.} is not adequate for such systems, we demonstrate the issue of indiscriminately training ADs without considering the different vulnerability levels of patients to evasion attacks. 
we train a \textit{k}NN-based AD using data from all 12 OhioT1DM patients. Figure \ref{Figure: Plot_FN_TP_kNN} shows sample CGM traces from patients A\_5 and A\_2 with adversarial samples generated using URET. The black and red horizontal lines indicate the maximum normal glucose values in fasting (125 mg/dL) and postprandial (180 mg/dL) states, respectively. Green dots mark adversarial glucose values that the AD correctly flags (i.e., TPs), whereas red dots mark missed adversarial values (i.e., FNs). Indiscriminate training produces lower recall for some patients (e.g., patient A\_2) than for others (e.g., patient A\_5). In particular, the AD misses many adversarial samples for patient A\_2 because of its poorly defined benign decision boundary, resulting in a higher FN rate. In addition, patient A\_2's CGM trace is naturally shifted toward higher values than that of patient A\_5, with glucose levels exceeding 130 mg/dL even without an attack. The AD is therefore more likely to label an adversarial sample as benign, interpreting the deviation as normal physiological variation rather than as an attack.

%\karthik{Can we say why this may be the case - what's happening in patient 2 specifically?}
% offers inequitable protection for the two patients since it flagged a higher percentage of adversarial samples from patient 5 than from patient 2. 
% \karthik{There are many issues with the above paragraph. What do you mean by not adequate? Also, as I have told you before, don't use "inequitable" protection as it suggests lack of fairness, which is not your focus. I still don't understand why flagging more samples from patient 5 is a problem. }
% \gargi{It's good that the graphs show FN, which is consistent with intro. }

\section{Methodology}
\label{Section: Methodology}
% \gargi{We already established that the paper is not just about risk profiling, but about a novel training strategy where we control the amount of anomalies in the training dataset. Risk profiling framework enables us to do that. I feel this is not coming out in the current structure of this section. I'd suggest splitting this section into three subsections - risk-profiling framework (this will be the most important subsection), noise removal, and introduction of anomalies (alternatively, you can combine these two under the same subsection called AD enhancement and make two headings). AD enhancement should not be a part of the risk profiling framework IMO. Steps 1-4 in Figure 3 belong to the risk profiling framework. To summarize, modify the introduction of the section and restructure the subsections. I would also recommend changing the caption of Fig. 3.}

\begin{figure}[t]
	\centering
	\includegraphics[scale=0.35]{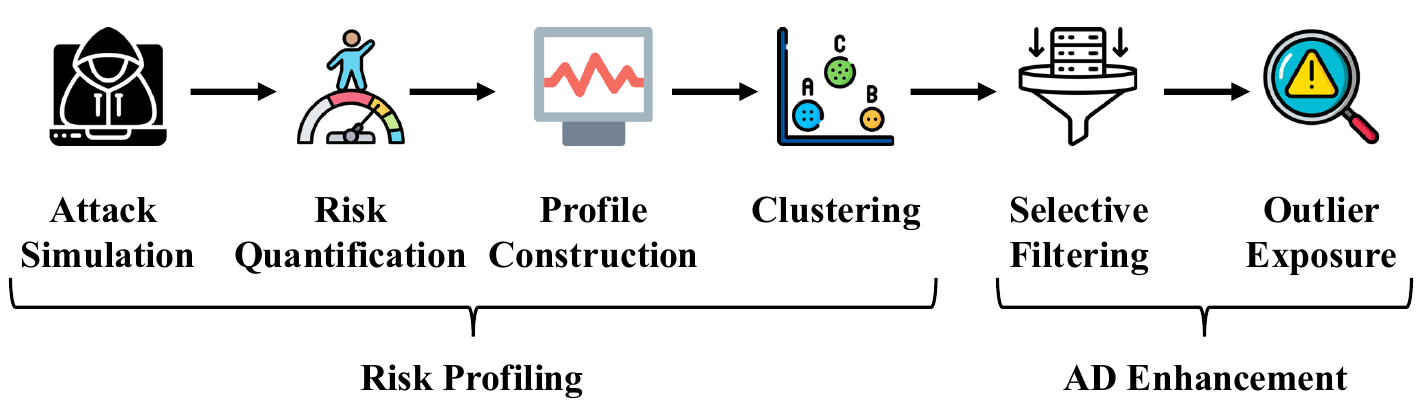}
	\caption[justification=centering]{The six steps of our proposed \systemname technique.}
	
	\label{Figure: Framework}
\end{figure}

% In this section, we introduce \systemname, our technique 
% our risk-aware outlier-exposed selective training strategy 
% to improve the recall of ADs against evasion attacks in healthcare. 
\systemname consists of two stages, risk profiling and AD enhancement, which together constitute a six-step process as shown in Figure \ref{Figure: Framework}. 
    % Risk profiling consists of steps 1--4, while steps 5--6 belong to AD enhancement. 
    We explain the six steps in this section.  
% \systemname is an iterative framework that is run periodically to ensure patient risk profiles are up to date. 
%	process that can be repeated from steps 1 through 5 whenever the patient or attacker behavior changes, 

%However, in our evaluation, we consider only a single iteration of \systemname before AD deployment. 

%	\karthik{Added this - please check} }
%\karthik{We should emphasize that this is an iterative process in that you can go through the loop multiple times.}
% \karthik{How do the steps map to the stages? Also, can we give the framework a name so we can refer to it consistently - right now, we say risk-aware, risk profiling, risk-based etc. for the same thing, which is confusing.}
% \gargi{Nitpick: In Fig 3, Step 5, can you change `defence' to AD?}

\subsection{Risk Profiling}
\label{Subsection: Risk_Profiling}
To determine the most suitable samples for AD training, the risk-profiling stage measures vulnerability before training or querying an AD. 
% We summarize each patient's attack-induced risk profile using the patient-level impact score defined after the instantaneous-impact formulas below. This definition depends only on the victim-model attack and its generated risk profile, not on an AD decision boundary or downstream detection results.
The risk profiling stage consists of four steps, as shown in Figure \ref{Figure: Framework}: (1) attack simulation, (2) risk quantification, (3) profile construction, and (4) clustering. 

% \karthik{I know we've used this figure before, but I wonder if it can be drawn better. For example, why is it a loop? Does this mean we go back to the first step after the fourth?}
% It's because it's an iterative process that never stops. We need to regularly do that to ensure the AD is up-to-date with the evolving threat landscape.

% The first four steps of Figure \ref{Figure: Framework} correspond to the risk profiling stage. It simulates the evasion attack, quantifies the risk imposed at each timestamp, converts those risks into patient-specific time-series profiles, and clusters the profiles into vulnerability groups. The key novelty of the risk profiling stage is assessing the temporal adversarial impact at the patient level rather than only at the aggregate model level. To elaborate on the operation of the proposed risk profiling stage, we take a closer look at its various steps. 

% \textit{Fifth}, it incorporates clustering insights by selectively training ADs on the less vulnerable patients to learn robust features that improve resilience against evasion attacks.

% \karthik{Can we say what is new or innovative about this framework? Also, why don't we talk about step 5 here?}
% step 5 belongs to AD enhancement and not risk profiling. I tried to make this clearer.

\textbf{Attack Simulation.} 
    % Our risk profiling stage is attack-agnostic, eliminating any dependency on the specifics of the attack algorithm. 
We simulate black-box attacks using URET~\cite{eykholturet} and white-box attacks using FGSM~\cite{goodfellow2015explaining}, PGD~\cite{mkadry2017towards}, and C\&W~\cite{carlini2017towards}.
We chose URET as a black-box baseline because it represents an adversary with limited model knowledge. 
% URET allows us to define customized input transformations and feature dependencies, which is important in healthcare, where attacks must respect temporal structure and domain-specific plausibility constraints. Although URET was developed for individual data points, healthcare data are often time series. We therefore extended URET to time-series data by flattening features across a fixed number of timestamps, allowing the adversary to manipulate the current timestamp while conditioning on the previous ones.
% the previous \textit{n} timestamps before manipulation. 
% Flattening preserves temporal order through a fixed positional structure. URET also supports dependencies among features in the flattened vector, allowing constraints on perturbations at neighboring timestamps.
We chose FGSM, PGD, and C\&W to represent full-knowledge adversaries with complementary optimization strategies. FGSM applies a single gradient-sign step~\cite{goodfellow2015explaining}, PGD iteratively applies and projects such steps within a bounded region~\cite{mkadry2017towards}, and C\&W jointly optimizes perturbation magnitude and attack success~\cite{carlini2017towards}.
% \karthik{You should add 1 sentence on how FGSM actually works.}
Evaluating \systemname against these attacks spans threat models from limited knowledge to complete model transparency, strengthening the practical applicability of our framework.

%\karthik{This should go much earlier.}
% 
% \karthik{Is there something special about URET that we should make it part of our framework? Why not say we use a framework for mounting evasion attacks, and later talk about how we extended URET in the experimental evaluation?}

%%%%%%%%%%%%%%%%%%%%%%%%%%%%%%%%%%%%%%%%%%%%%%%%%%%%%%%%%%%%

\textbf{Risk Quantification and Profile Construction.} To quantify attack-induced impact at each timestamp, we devise a proxy score that combines (a) the magnitude of the deviation between the benign and adversarial samples and (b) a nonnegative empirical influence weight for each manipulated variable. The magnitude captures the size of the perturbation, while the weight captures the strength---but not the direction---of the variable's association with the modeled outcome. The resulting score is an empirical adversarial-impact proxy rather than a clinically calibrated measure.
% \karthik{Be more concrete}
% For example, transitioning a diabetic patient from hypoglycemia to hyperglycemia is more life-threatening than from normal glucose to hyperglycemia.

We propose the generic weighted-sum formula in Equation \ref{Equation: Instantaneous_Error} to calculate the instantaneous attack-induced impact of manipulating a patient's vital measurements. Here, $R(t)$ denotes the instantaneous impact at time $t$, and $Z_i(t)$ represents the $i$-th impact factor. The index $i \in \{1,2,\ldots,n\}$ enumerates the impact factors, and $n$ is their total number. Each factor captures a specific type of manipulation; for example, $Z_1(t)$ may correspond to the difference between benign and adversarial glucose values, while $Z_2(t)$ may represent manipulated blood pressure. This formulation allows \systemname to support broader adversarial capabilities rather than being tied to the capability model considered in this paper.
%\karthik{You haven't introduced the model yet - you need to do that first.}

% \karthik{Can we explain why this makes sense intuitively? What does the patient data have that we leverage?}

\begin{equation}
	\label{Equation: Instantaneous_Error}
    R(t) = \sum_{i=1}^{n} S_i \cdot Z_i(t), \qquad t \in \mathbb{N}
\end{equation}
% \gargi{What's i and n?}

The parameter $S_i$ is the nonnegative empirical influence weight associated with the $i$-th impact factor.
$S_i$ captures the strength of the association between the manipulated variable and the modeled outcome. For example, in a mortality prediction model, heart rate may have a stronger association with mortality than blood pressure. The weight $S_i$ would therefore be larger for heart rate than for blood pressure.
To estimate these weights, we fit a linear regression model for regression tasks (OhioT1DM) or a logistic regression model for classification tasks (MIMIC and PhysioNet CinC)~\cite{mehta2016regression}, obtaining a signed coefficient $\beta_i$ that corresponds to each feature. Its sign represents the direction of the conditional association with the encoded model outcome; it does not represent positive or negative clinical severity. We therefore define $S_i=|\beta_i|$ so that the impact contribution is nonnegative. If multiple features are manipulated, the predictors should first be standardized and their magnitudes normalized as $S_i=|\beta_i^{\mathrm{std}}|/\sum_j|\beta_j^{\mathrm{std}}|$ to prevent feature scale from determining the weights.
% \karthik{Cite papers on this}
This approach estimates each variable's empirical association with the modeled outcome and provides a data-driven weighting scheme for the impact factors. Clinically validated severity weights would require domain-expert input and are outside the scope of this work.

% We derive the instantaneous impact formula for each of the three datasets from Equation \ref{Equation: Instantaneous_Error}. 
We define the impact factor $Z_i(t)$ as the squared difference between the manipulated vital-sign value and its original value. This squared difference, inspired by the mean squared error, assigns larger proxy scores to larger deviations.
Since we manipulate one physiological feature per dataset in our experiments,
%	\karthik{We need an adversary model early on} 
our impact formulas contain a single impact factor. For a multi-feature perturbation, Equation \ref{Equation: Instantaneous_Error} would contain one impact factor for each manipulated feature.
For patient $p$, let $R_p(t)$ denote the instantaneous impact at timestamp $t$, 
% computed using the corresponding dataset-specific equation above,
and let $T_p$ denote the number of timestamps in the patient's impact profile. Because $S_i$ and $Z_i(t)$ are nonnegative, $R_p(t)\geq 0$. We define the patient-level impact score as $I_p=\frac{1}{T_p}\sum_{t=1}^{T_p}R_p(t)$. A larger $I_p$ indicates that the configured attack produces a larger empirically weighted perturbation under our operational definition; it does not imply greater clinical instability.

%%%%%%%%%%%%%%%%%%%%%%%%%%%%%%%%%%%%%%%%%%%%%%%%%%%%%%%%%%%%%%%%

\textbf{Clustering.} After generating the patients' risk profiles, the framework first clusters their temporal shapes without assigning vulnerability semantics. It then applies the same deterministic labeling rule in every experiment: the cluster with the lower median $I_p$ is labeled ``less vulnerable,'' and the other cluster is labeled ``more vulnerable.''
% Thus, neither attack-success percentages nor AD performance are used to construct or label the clusters; these outcomes are reserved for subsequent validation.
%\karthik{Point to the section where we discuss the similarity score}. 
%\nawawy{We do not discuss similarity score. It is common knowledge of clustering algorithms where they typically use euclidean distance, Manhattan distance, cosine similarity, etc. to measure similarity.}

% \systemname is clustering-agnostic because its vulnerability-scoring and selective-training components are independent of the clustering algorithm. We selected hierarchical clustering for three reasons. \textit{First}, partitioning algorithms such as k-means require the number of clusters a priori and assume approximately spherical clusters, assumptions that may not reflect the structure of physiological data. Hierarchical clustering does not impose either requirement and can group patients into homogeneous groups~\cite{Sarstedt2014}. \textit{Second}, density-based algorithms such as DBSCAN are sensitive to density parameters and may leave some patients unclustered~\cite{lai2019new}. \textit{Third}, hierarchical clustering produces a transparent dendrogram that visualizes patients with similar physiological characteristics~\cite{johnson1967hierarchical}.

\systemname's vulnerability scoring and labeling rules are separable from the underlying clustering algorithm. However, we chose agglomerative hierarchical clustering because ROAST represents patients as temporal attack-impact risk profiles and defines similarity between complete profiles rather than between each profile and an estimated centroid. Pairwise hierarchical clustering preserves this representation without requiring profiles to be reduced to fixed-dimensional vectors or summarized by synthetic temporal prototypes. In addition, the resulting hierarchy provides an auditable account of patient grouping~\cite{johnson1967hierarchical}. 

We use complete linkage and cut the resulting dendrogram into two clusters to support ROAST's binary patient-selection decision. We then use cluster-level median impact scores $I_p$ to assign vulnerability labels to each cluster such that the cluster with the lower $I_p$ is labeled ``less vulnerable,'' and the other cluster is labeled ``more vulnerable.''

%\karthik{Can we explain the rationale for clustering? Why do it?}
% \karthik{I am confused. Is clustering algorithim part of the framework? If so, then why is the framework independent of the algorithm? Did you mean it can support different algorithms?}
% Yes. I rephrased to make it clearer.

%\karthik{Can we give an example of the steps here?}

% \gargi{In our experiments, we are always using two clusters. We need to justify that.}
% I will justify this in the evaluation section. The main reason why we use two clusters is that it yields the largest vertical difference between nodes in the tree hierarchy.
%%%%%%%%%%%%%%%%%%%%%%%%%%%%%%%%%%%%%%%%%%%%%%%%%%%%%%%%%%%%%%%%
\subsection{Anomaly Detector Enhancement}
\label{Subsection: Anomaly_Detector_Enhancement}
\begin{figure*}[t]
	\centering
	\includegraphics[width=0.9\textwidth]{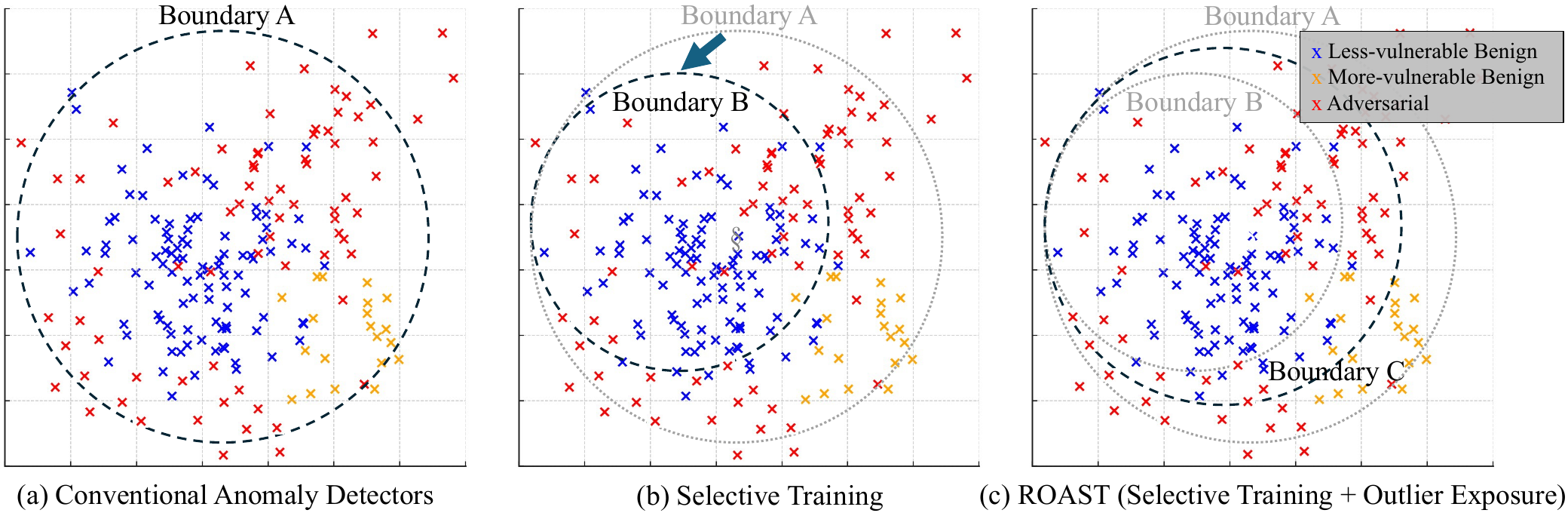}
    \caption{Illustrative AD training strategies: (a) indiscriminate training on benign data produces a broad boundary (Boundary A) that misses attacks, (b) selective training on less-vulnerable patients tightens the boundary (Boundary B), improving recall but potentially increasing false positives, and (c) \systemname adds outlier exposure to learn Boundary C, improving recall while largely preserving precision.}
	\label{Figure: AD_Training}
	
\end{figure*}

Figure \ref{Figure: AD_Training} illustrates the intuition behind \systemname. In conventional AD training (Figure \ref{Figure: AD_Training}a), benign data from all patients are used indiscriminately. Motivated by the case study, we hypothesize that cross-patient heterogeneity can broaden the learned benign region (Boundary A) and lower recall. We therefore propose a two-step solution that enhances recall while largely preserving precision: (a) selective filtering and (b) outlier exposure.

\textbf{Selective Filtering.} We use the clusters obtained in §\ref{Subsection: Risk_Profiling} to retain the patients with lower impact scores for AD training. 
% The case study motivates, but does not establish, an association between attack-induced impact and benign-trace heterogeneity; 
Following the hypothesis developed in §\ref{Section: Motivational_Case_Study}, our evaluation tests whether this patient-level selection improves AD performance. \systemname does not classify clinically abnormal observations as errors or remove individual measurements based on clinical reference ranges.

% \karthik{Sorry, this is confusing. We just talked about risk profiling etc. to cluster the patients. I am now expecting you to tell me how to do the training of the ADs using the clusters. This establishes the rationale for the entire framework, and should go in the beginning of this section instead. }
% \gargi{Establish the connection with Section 4.1}

\textbf{Outlier Exposure.} With conventional selective training (Figure \ref{Figure: AD_Training}b), the AD is trained only on benign data from less vulnerable patients. This narrows the learned benign region from Boundary A to Boundary B, improving recall but potentially increasing FPs.
For this reason, we augment selective training with outlier exposure (Figure \ref{Figure: AD_Training}c), which injects controlled adversarial samples from less vulnerable patients during training. These plausible, near-boundary anomalies ``blend in'' with normal data, giving the AD a richer representation of the sample space and a more precise decision boundary than benign-only selective training. % \karthik{An example would really help here - you can perhaps explain what these adverserial samples mean in the BG prediction.}
% \gargi{Could you clarify this sentence? Not sure how adversarial examples blending in results in a richer representation of benign data - this sounds contradictory.}
This expands the boundary to Boundary C, improving recall over indiscriminate training while minimizing the impact on precision from benign-only selective training. The injected samples are structured anomalies near the benign decision boundary rather than arbitrary outliers~\cite{paschali2018generalizability}.
We evaluate the effectiveness of this outlier exposure strategy in §\ref{Subsection: Experimental_Results} (RQ2 answer).
% \karthik{Do we evaluate how effective this outlier exposure is? If so, point to the section in the results where we do so.}

% \karthik{This argument seems hand-wavy to me. Do we have evidence to back it up? Or at the least, we need citations.}
% \gargi{Need to justify why we select this strategy - Is there any characteristic feature of these anomalous points? Why do you think this is just the right amount of outliers?}

% \begin{figure*}[t]
%     \centering
%     \includegraphics[width=0.9\textwidth]{Figures/AD_Training.pdf}
%     \caption{Anomaly Detector Training Strategies.}
%     \label{Figure: AD_Training}
%     
% \end{figure*}

\section{Evaluation}
\label{Section: Evaluation}

This section answers
% ask\gargi{we answer the following questions with experiments, right?} 
the following research questions:
\begin{enumerate}[label = {\textbf{RQ\arabic*:}}, align=left]
    \item Does indiscriminate AD training produce low recall (i.e., a high FN rate)? If so, why?
    % \gargi{Lower compared to what?} 
    % {\color{red} 
    %in the healthcare domain
    % }
    % when {\color{why}}?
    \item How much does \systemname improve AD recall?
    % the selective training with outlier exposure strategy 
%    (i.e., lower FNs) 
    \item What is the corresponding impact on precision?
%    \karthik{Rephrased as a question rather than a conclusion.}
    % \gargi{If we are only talking about recall, then there's just one strategy (filtering out the most vulnerable patients), right? Shouldn't we mention keeping precision intact too}?
    \item How much does \systemname reduce the training set size and training time?
%    \karthik{Added time and rephrased}
	% \item How sensitive is \systemname's clustering to variations in regression-derived influence weights and dendrogram cut thresholds?
	\item Does \systemname's patient selection generalize across attack algorithms, or is it tied to the attack used for risk profiling?
\end{enumerate}

RQ2--RQ4 use three established baselines spanning local-density (\textit{k}NN), one-class boundary (One-Class SVM), and temporal generative detection (MAD-GAN). Deep SVDD and LSTM-AE serve as architectural robustness checks, extending the fairness and cross-attack analyses to neural one-class and reconstruction-based detectors (RQ5).

% We use a two-tier evaluation scope. The primary RQ2--RQ4 experiments evaluate \textit{k}NN, One-Class SVM, and MAD-GAN on all three datasets and provide the headline recall, precision, and overhead results. The fairness and cross-attack robustness analyses additionally include Deep SVDD and LSTM-AE, yielding five ADs.

\subsection{Experimental Setup}
\label{SubSection: Experimental_Setup}
\textbf{Victim models.} Since DNN prediction models in healthcare are often confidential, 
%~\cite{dreamed}
%~\cite{dreamed, schertz2023sepsis, sepsis_immunoscore}, 
we approximate victim models using time-series prediction models from prior work: (1) Rubin-Falcone et al. for blood glucose prediction~\cite{rubin2020deep}, (2) Gupta et al. for mortality prediction~\cite{gupta2022extensive}, and (3) Nejedly et al. for sepsis prediction~\cite{nejedly2019prediction}. The glucose prediction model achieved average root-mean-square errors of 18.2 and 31.7 and average mean absolute errors of 12.8 and 23.6 at the 30- and 60-minute horizons, respectively~\cite{rubin2020deep}. On ICU data, the mortality prediction model achieved an AUROC of 0.87 and an AUPRC of 0.49~\cite{gupta2022extensive}. The sepsis prediction model achieved a normalized utility score of 0.278 under the 2019 PhysioNet/Computing
in Cardiology Challenge utility function~\cite{nejedly2019prediction}.
%\karthik{Are these numbers good or bad compared to commercial models?}
%\nawawy{It is really difficult to tell because each model uses its own perfoemance metrics, and it's also difficult to obtain similar data for commercial models.}
%\karthik{Can we say how good was the baseline accuracy in the absence of attacks?}

\textbf{Attack algorithms.} To evaluate \systemname under complementary threat models, we 
% employ (1) 
use URET~\cite{eykholturet}
% , a black-box, non-gradient-based framework, 
% and (2) 
and FGSM~\cite{goodfellow2015explaining}
% , a white-box, gradient-based method. We use these attacks 
for the main recall (RQ2), precision (RQ3), and overhead (RQ4) experiments. For fairness across clusters (RQ2) and cross-attack generalization analysis (RQ5), we additionally include two multi-step white-box attacks, PGD~\cite{mkadry2017towards} and C\&W~\cite{carlini2017towards}, to test whether \systemname's patient selection remains effective across clusters and against stronger attacks not used during risk profiling.

% For OhioT1DM, we manipulate current CGM measurements to cause errors in future CGM predictions. 
For OhioT1DM, the adversary raises current CGM measurements to bias the DNN toward higher future CGM predictions.
% \gargi{For these attacks, would the attacker know when in future the misclassification is going to happen?} 
% The adversary may not know the exact time at which the effect of manipulations will be reflected on future predictions. But the rationale is: raise the current CGM values and the DNN thinks that it is safe to predict higher values in the future, causing mispredictions. I added a sentence above to clarify this.
% ME
Manipulated values are constrained to remain physiologically plausible: 125--499 mg/dL or 180--499 mg/dL for OhioT1DM fasting and postprandial attacks, respectively (499 mg/dL is the highest reported glucose level in OhioT1DM), 150--200 bpm for MIMIC pacemaker rates, and 120--140 bpm for PhysioNet CinC heart rates. These bounds ensure that the generated attacks remain clinically plausible rather than relying on obviously unrealistic perturbations, which would be easily detected. For example, the 150--200 bpm pacemaker rate range for the MIMIC dataset was selected to represent the zone immediately above the highest recorded value (130 bpm) in the dataset. Per the threat model (§\ref{Section: Threat Model}), the adversary reads correlated features via the compromised smartphone to preserve physiological coherence across features.

\textbf{Attack hyperparameters.} We use a fixed configuration for each attack throughout the experiments. (1) \textit{URET:} We use model-loss-guided beam search with \texttt{search size=5}, \texttt{maximum depth=2}, and a brute-force ranker. (2) \textit{FGSM:} We use a single $L_\infty$ gradient-sign step with a perturbation budget of $\epsilon=20$ mg/dL for OhioT1DM, $\epsilon=0.2$ in the MIMIC model's input representation, and $\epsilon=0.02$ after feature-wise normalization for PhysioNet CinC. (3) \textit{PGD:} We use the $L_\infty$ norm and \texttt{iterations=10}. The $(\epsilon,\alpha)$ budget and step-size pairs are $(20,4)$ mg/dL for OhioT1DM, $(0.2,0.04)$ for MIMIC, and $(0.02,0.004)$ in normalized units for PhysioNet CinC. (4) \textit{C\&W:} We jointly optimize the mean-squared $L_2$ perturbation penalty and task loss, using a trade-off constant of $c=1$, Adam with \texttt{iterations=50}, and no early stopping. The Adam learning rates are $1.0$, $0.05$, and $0.01$ for OhioT1DM, MIMIC, and PhysioNet CinC, respectively. OhioT1DM uses mean-squared prediction loss, whereas MIMIC and PhysioNet CinC use binary cross-entropy and cross-entropy, respectively.

\textbf{Clustering setup.}
We cluster risk profiles hierarchically using the DTAIDistance library, with pairwise distances computed through dynamic time warping (DTW) and complete linkage. DTW permits nonlinear alignment and thus accounts for temporal misalignment in physiological data. We do not use Euclidean distance because it assumes pointwise alignment and equal-length sequences, which can misrepresent the similarity of temporally shifted clinical signals. Complete linkage merges clusters according to the maximum inter-cluster distance, producing compact, well-separated clusters and reducing the chance that high-impact patients enter the ``less vulnerable'' group.
%Empirically, complete linkage yielded tighter separation of adversarial risk profiles than average linkage.

\textbf{AD hyperparameters.}
We did not perform AD-specific hyperparameter tuning because our goal is to isolate the effect of \systemname by comparing it with the baselines under fixed AD settings. We therefore use one fixed configuration for each detector across all experiments and training strategies. We use the following AD hyperparameters.  
(1) \textit{kNN:} We use the PyOD library with \texttt{neighbors=7} and \texttt{contamination=0.5}. (2) \textit{One-Class SVM:} We use the scikit-learn library with \texttt{kernel=rbf}, \texttt{gamma=auto}, \texttt{tol=0.001}, and \texttt{nu=0.5}.
(3) \textit{MAD-GAN:} We follow the configuration in \cite{li2019mad}, with \texttt{epochs=100}, \texttt{learning rate=0.1}, \texttt{batch size=500}, \texttt{sequence length=12}, and \texttt{sequence step=1}. (4) \textit{Deep SVDD:} We use the PyOD implementation with two hidden layers containing 64 and 32 neurons, ReLU hidden activations, Adam optimization, \texttt{epochs=100}, \texttt{batch size=32}, \texttt{dropout=0.2}, an $L_2$ regularization \texttt{coefficient=0.1}, and \texttt{contamination=0.5}. The hypersphere center is initialized from the network's first forward pass, and the contamination setting places the anomaly threshold at the median training decision score. (5) \textit{LSTM-AE:} We use windows of 10 consecutive observations and a single-layer LSTM encoder--decoder with \texttt{hidden units=32} and a linear reconstruction head. We train using mean-squared reconstruction error and Adam with \texttt{learning rate=$10^{-3}$}, \texttt{epochs=20}, and \texttt{batch size=256}. The anomaly score is the mean reconstruction error over all timestamps and features in a window; the threshold is the median training reconstruction error (i.e., the 50th percentile, corresponding to \texttt{contamination=0.5}). Although \texttt{contamination=0.5} and \texttt{nu=0.5} reflect the 1:1 benign-to-adversarial composition of the OE arms, we retain them for the benign-only baseline to isolate the effects of patient selection and OE rather than changing detector hyperparameters between strategies. 
% \karthik{We need to justify why these particular hyper-parameters were chosen. Are these based on prior work? If so, cite them.}
% \gargi{This is not a part of experimental setup, right? Should go to Sec 5.2.1.} 
% Please check Karhtik's comment below this paragraph.
% ME
%We chose hierarchical clustering to group patients into vulnerability clusters for three reasons
%%.
%~\cite{Noble_2024}. 
%\textit{First}, we do not need to specify the number of clusters in advance, as it is difficult to know the optimal number of clusters a priori. Instead, the dendrogram can be pruned at the desired level based on the inter-cluster distances. \textit{Second}, the dendrogram helps visually observe patients with similar physiological characteristics at different levels of the hierarchy. \textit{Third}, it is suitable for clinical research as it categorizes mixed populations into homogeneous groups. 
%\karthik{Need citations for these claims.}
% \karthik{I don't know enough about other clustering algorithms to appreciate this, but if this is an implementation choice, perhaps we can move it to the experimental setup instead.}

\textbf{Hardware platform.} 
All experiments were performed on a workstation running Ubuntu 20.04 LTS with AMD Ryzen Threadripper 3960X (24 cores, 2.2 GHz), 64 GB RAM, and dual NVIDIA GeForce RTX 3090 GPUs (24 GB each).
%\karthik{Move this to the experimental setup.}

\subsection{Experimental Results}
\label{Subsection: Experimental_Results}
\subsubsection{Risk Profiling}
We first present the results of the individual steps and then answer the RQs.
% \karthik{Is this accurate?} 

\begin{figure}[t]
	\centering
	\includegraphics[width=0.45\textwidth]{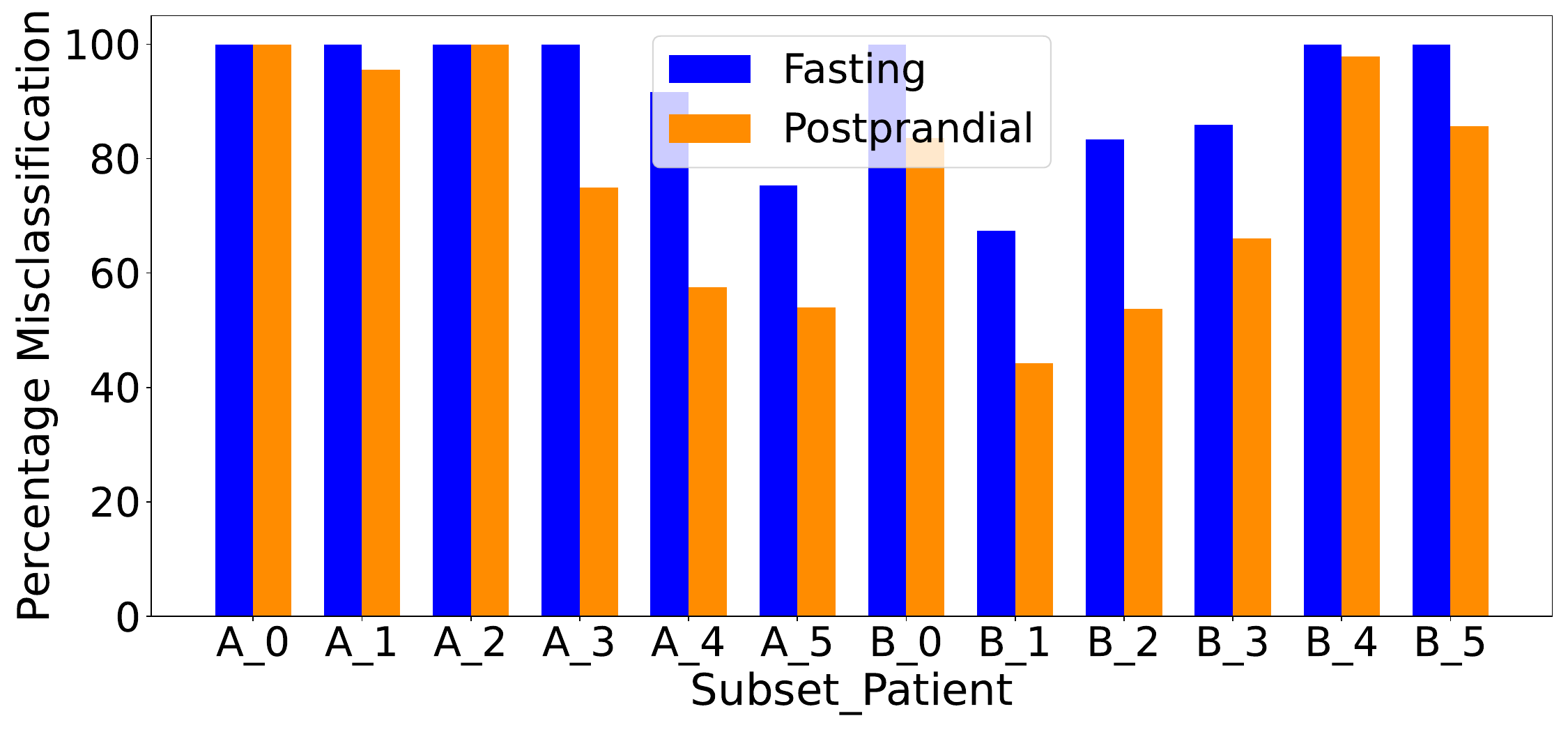}
	\caption[Percentage of Normal Glucose Mispredictions]{OhioT1DM normal glucose samples misclassified as hyperglycemic under fasting and postprandial attacks.}
	\label{Figure: Normal_to_Hyper}
	
\end{figure}

\hspace*{\parindent}\textbf{Attack Simulation Results.} We define a successful evasion attack as one that transitions a patient from a safe state to an unsafe state. To demonstrate this, we launched the attack described in §\ref{SubSection: Experimental_Setup} on all three datasets. For example, on OhioT1DM, we tested the attack on all 12 patients 
% \karthik{What about the other datasets? Is this only for the Ohio dataset?}
%\karthik{Is this the same attack we launched earlier in Section 3?}
using two attack scenarios, fasting (i.e., $>$125 mg/dL) and postprandial (i.e., $>$180 mg/dL). Figure \ref{Figure: Normal_to_Hyper} shows the percentage of originally normal glucose samples that were misdiagnosed as hyperglycemic. The results show that some patients are more vulnerable to the attack than others. For example, patients A\_0 and A\_2 are almost always vulnerable to the attack regardless of the attack scenario, while patients B\_1 and A\_5 appear to be the least vulnerable to the attack. 
This patient-level variation complements the earlier motivational example: the present experiment identifies A\_2 as more attack-vulnerable than A\_5, and Figure \ref{Figure: Plot_FN_TP_kNN} shows that the AD missed more adversarial samples for A\_2 than for A\_5.
% \gargi{What is subset A and B? The last time we mentioned them was in   2. There's a long gap after that and the reader would forget by the time they reach Sec 5.} 
%\karthik{Can we relate this to our earlier findings in Section 3. Are these the same patients? Also, why do you call them A and B here?}

% The results also show that it is usually easier to attack a patient in a fasting state.

% \gargi{Not sure why this information is important here. Can we tie the outcome of this step to the next step? Also, we should add the results for this step for the other datasets too.}
% I removed unnecessary information about fasting/postprandial. I am already connecting the results of the clustering step to this figure when I talk about the obtained clusters and how they match with the percentage misclassifications. However, the only issue is that I cannot show similar results on the other two datasets since they contain approximately. 13k and 40k patients, so it is impractical to show in a graph; that is why I am showing a summary of the number of patients in each cluster in \ref{Table: Summary_of_Clusters}.

\textbf{Risk Quantification and Profile Construction.} 
% To estimate the influence weights $S$ 
% in Equation \ref{Equation: Instantaneous_Error}
% for each of the three tasks, we fit a linear regression model to OhioT1DM and logistic regression models to MIMIC and PhysioNet CinC.
Table \ref{Table: Severity_Coefficients} reports both the signed regression coefficient $\beta$ and the nonnegative weight $S=|\beta|$ for each manipulated feature. All three fitted coefficients are negative, indicating inverse conditional associations under the encoded outcomes. For each patient, we preserve the temporal order of the instantaneous impact scores to form the risk profile $[R(1), R(2), \ldots, R(T)]$, where $T$ is the number of timestamps.
% \karthik{Combine them how?}

% Please add the following required packages to your document preamble:
% \usepackage{graphicx}
\begin{table}[t]
\centering
\caption{Signed regression coefficients and nonnegative empirical influence weights for the manipulated features.}
\label{Table: Severity_Coefficients}
\resizebox{\columnwidth}{!}{%
\begin{tabular}{|l|l|l|l|}
\hline
\textbf{Dataset} & \textbf{Feature} & \textbf{$\beta$} & \textbf{$S=|\beta|$} \\ \hline
OhioT1DM       & CGM & -10.78 & 10.78 \\ \hline
MIMIC          & PMR & -0.76  & 0.76  \\ \hline
PhysioNet CinC & HR  & -4.52  & 4.52  \\ \hline
\end{tabular}%
}
\end{table}

\textbf{Clustering.} 
To group patients into different clusters based on their vulnerability to the attack, we use the time-series risk profiles obtained from the risk quantification step.
We focus the detailed patient-level clustering discussion on OhioT1DM because its small cohort permits interpretable per-patient analysis and direct comparison with Figure \ref{Figure: Normal_to_Hyper}. Because MIMIC and PhysioNet CinC are much larger, we summarize their clustering outcomes in Table \ref{Table: Summary_of_Clusters}.
%\textbf{Clustering.}
% \gargi{At the beginning of each paragraph, say how you are using the results of the previous paragraph here, in a few words.} 
% Figure \ref{Figure: Clusters} shows the time-series risk profiles for each of the 12 patients of the OhioT1DM dataset. It also shows the resulting dendrogram from hierarchically clustering the 12 patients and the distance between clusters at each level. 

We split the patients into two clusters. Patients A\_0, A\_1, A\_2, A\_3, A\_4, B\_0, B\_3, B\_4, and B\_5 form one cluster; patients A\_5, B\_1, and B\_2 form the other. Under the predefined rule from §\ref{Subsection: Risk_Profiling}, the latter cluster has the lower median impact score and is therefore labeled ``less vulnerable.'' The misclassification percentages in Figure \ref{Figure: Normal_to_Hyper} provide a post-hoc validation, showing that representative members A\_5 and B\_1 also have comparatively low attack-success rates.
\begin{table}[t]
\centering
\caption{Cluster sizes and normalized inter-cluster distances.}
\label{Table: Summary_of_Clusters}
\resizebox{\columnwidth}{!}{%
\begin{tabular}{|l|l|l|l|}
\hline
\textbf{Dataset} & \textbf{Less Vulnerable} & \textbf{More Vulnerable} & \textbf{Inter-cluster Distance} \\ \hline
OhioT1DM       & 3     & 9     & 32.46 \\ \hline
MIMIC          & 889   & 12111 & 37.84 \\ \hline
PhysioNet CinC & 11299 & 29037 & 26.10 \\ \hline
\end{tabular}%
 }
\end{table}

\begin{figure}[t]
	\centering
	\includegraphics[width=0.4\textwidth]{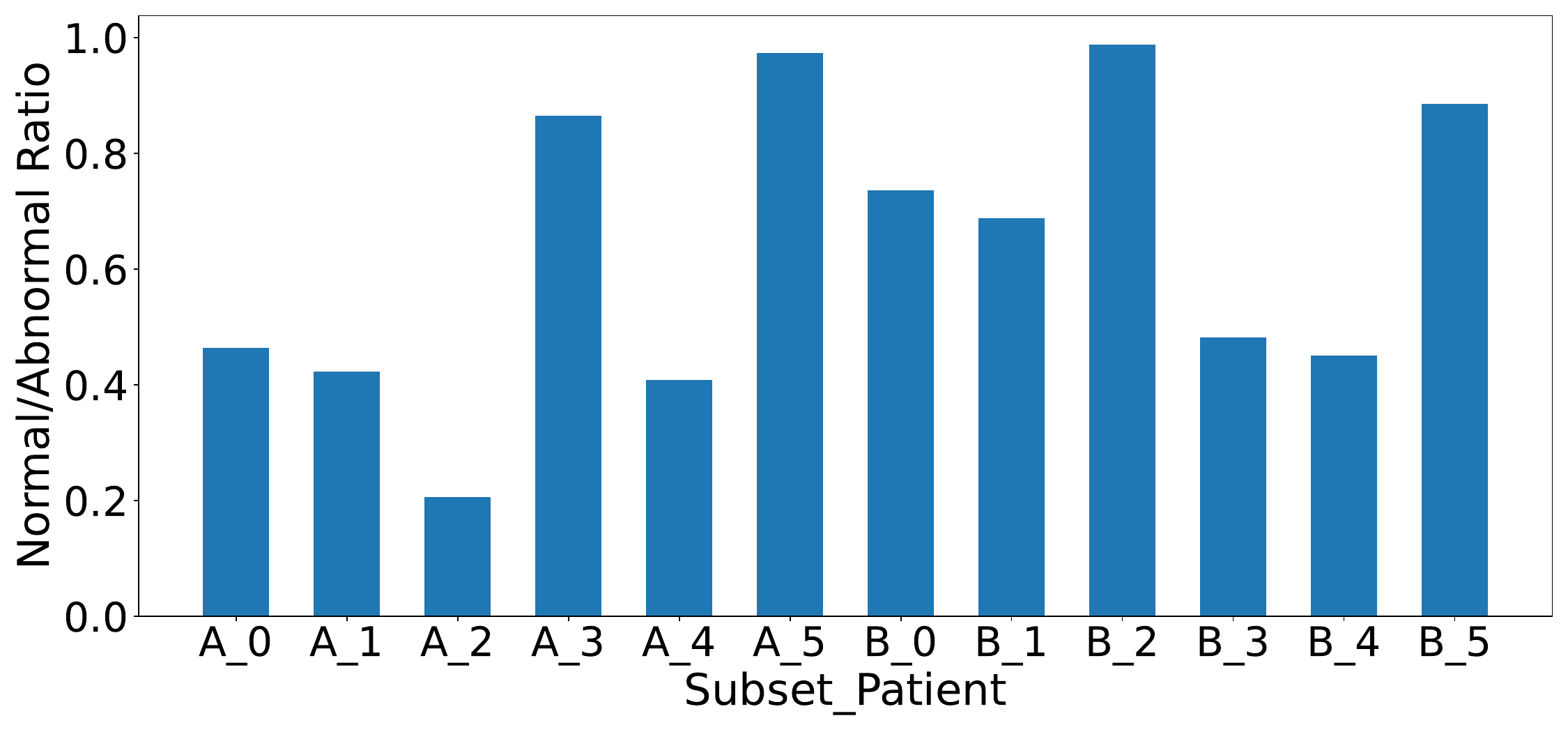}
	\caption[Ratio of Normal to Abnormal Instances]{Ratio of glucose measurements in the clinically normal range to those in abnormal ranges (i.e., hypoglycemia or hyperglycemia) in benign OhioT1DM traces.}
	\label{Figure: Ratio_Normal_To_Abnormal}
	
\end{figure}

\paragraph{\textbf{Answer to RQ1}}
To explain the FN gap caused by indiscriminate training, Figure \ref{Figure: Ratio_Normal_To_Abnormal} plots the benign normal-to-abnormal glucose ratio for the 12 OhioT1DM patients, where \emph{abnormal} refers specifically to clinically defined hypoglycemic or hyperglycemic glucose measurements. It does not denote measurement noise or invalid data. Patients A\_5 and B\_2, who belong to the less vulnerable cluster, show the highest ratios, making adversarial samples easier for the AD to distinguish from benign samples.
% (Figure \ref{Figure: Quadrants}). 
In contrast, patient A\_2 (more vulnerable cluster) shows the lowest ratio, indicating that abnormal samples are already common in the benign trace.
% (Figure \ref{Figure: Quadrants}). 
Adversarial samples may therefore be more likely to be misclassified as benign when they resemble variation already present in a patient's benign trace. This result is consistent with our hypothesis that training on less vulnerable patients reduces FNs, albeit with a possible impact on precision.
% \gargi{We should mention the number of clusters. cluster distance and number of patients in each cluster for the other datasets as well.}

\insightbox{RQ1 Insight}{Patients who are less vulnerable to attack tend to have a higher ratio of normal to hypo-/hyperglycemic samples in their benign data than more vulnerable patients.}
%\karthik{We should make this more general than just the glucose values as RQ1 is more general.}

\newlength{\oldheight}
\settoheight{\oldheight}{\includegraphics[width=0.5\columnwidth]{Figures/Ratio_Normal_To_Abnormal.pdf}}

% \begin{figure}[t]
%     \centering
%     \includegraphics[width=\columnwidth,height=\oldheight]{Figures/example.pdf}
%     \caption{Your caption here}
%     \label{fig:stretch}
% \end{figure}

% \begin{figure}[t]
% \centering
% \includegraphics[width=0.475\textwidth]{Figures/Quadrants.pdf}
% \caption[Four Quadrants of Blood Glucose Samples]{The four quadrants of glucose samples: (a) benign normal: normal glucose in absence of attack, (b) benign abnormal: high or low glucose in absence of attack, (c) malicious abnormal: samples intentionally manipulated to fall in the high or low glucose ranges, and (d) malicious normal: samples intentionally manipulated to fall in the normal glucose range.}
% \label{Figure: Quadrants}
% 
% \end{figure}

\subsubsection{Anomaly Detector Enhancement}
\paragraph{\textbf{Answer to RQ2}}
To evaluate \systemname and determine the potential recall improvement from selectively training ADs over indiscriminate training,
% most suitable selective training strategy (\textbf{RQ2})
% \gargi{I think we should slightly re-word RQ2. Before section 5, we have already established that training with least vulnerable subset is the best strategy indeed. At this point, the question shouldn't be what's the best strategy. The question should be about how much improvement we observe.}, 
% {\color{red}for each dataset,} 
we train the three primary ADs---\textit{k}NN, One-Class SVM, and MAD-GAN~\cite{li2019mad}---using
% {\color{red} \textbf{Suggested substitute for the rest of the paragraph:}} 
five training sets per dataset. ``All Patients (Benign)'' follows conventional indiscriminate training on benign data from the full cohort. ``All Patients (OE)'' trains on the full cohort but with outlier exposure, serving as an ablation that controls for OE: because switching from ``All Patients'' to ``Less Vulnerable (OE)'' changes two parameters simultaneously---the training set is reduced to the less vulnerable subset and OE is applied---``All Patients (OE)'' isolates the contribution of OE from that of patient selection. ``Less Vulnerable (OE)'' and ``More Vulnerable (OE)'' use the corresponding patient clusters, while ``Random Samples (OE)'' randomly draws a patient subset of the same size as the less-vulnerable group, and averages over 10 runs to reduce random effects. 
	
	We treat ``All Patients (Benign)'', ``All Patients (OE)'', and ``Random Samples (OE)'' as baseline strategies because they do not exploit the vulnerability information exposed by risk profiling. 
% and improve the accuracy of the results
% \gargi{What is meant by improving accuracy here?}. 
All selective subsets and ``All Patients (OE)'' employ outlier exposure by injecting adversarial samples into benign data. Comparing ``Less Vulnerable (OE)'' against ``All Patients (Benign)'' measures the total gain from \systemname. Comparing ``Less Vulnerable (OE)'' against ``All Patients (OE)'' attributes any residual gap to patient selection rather than to OE alone. Comparison with the other subsets further tests whether gains are specific to the less vulnerable cluster rather than a coincidental sampling effect. All ADs are tested on patients from both clusters, regardless of the training subset used, so the evaluation reflects detector behavior across the full patient population rather than only the selected training cluster.

% four different subsets of each dataset. The ``Less Vulnerable'' and ``More Vulnerable'' subsets contain data exclusively from their respective patient clusters. The ``Random Samples'' subset is constructed by randomly drawing patient data, repeated across 10 runs and averaged to reduce random errors and improve the accuracy of the results. This allows us to test whether the gains observed from training on less vulnerable patients arise purely coincidentally. Finally, to demonstrate the effect of selective training on less vulnerable instances, we train the detectors indiscriminately on the ``All Patients'' subset. We treat ``All Patients'' and ``Random Samples'' as baseline strategies since they do not incorporate risk profiling insights. All subsets employ outlier exposure by injecting adversarial samples into benign data, except for ``All Patients'', which relies solely on benign data. This serves as a direct comparison to the conventional AD training strategy commonly used in the literature.

\begin{figure*}[t]
	\centering
	\scalebox{0.9}{% scale the entire figure by 90%
		\begin{minipage}{\textwidth}
			\centering
			% First subfigure
			\subfloat[OhioT1DM\label{Figure: OhioT1DM_Recall}]{\begin{minipage}{0.42\textwidth}
				\centering
				\includegraphics[width=\linewidth]{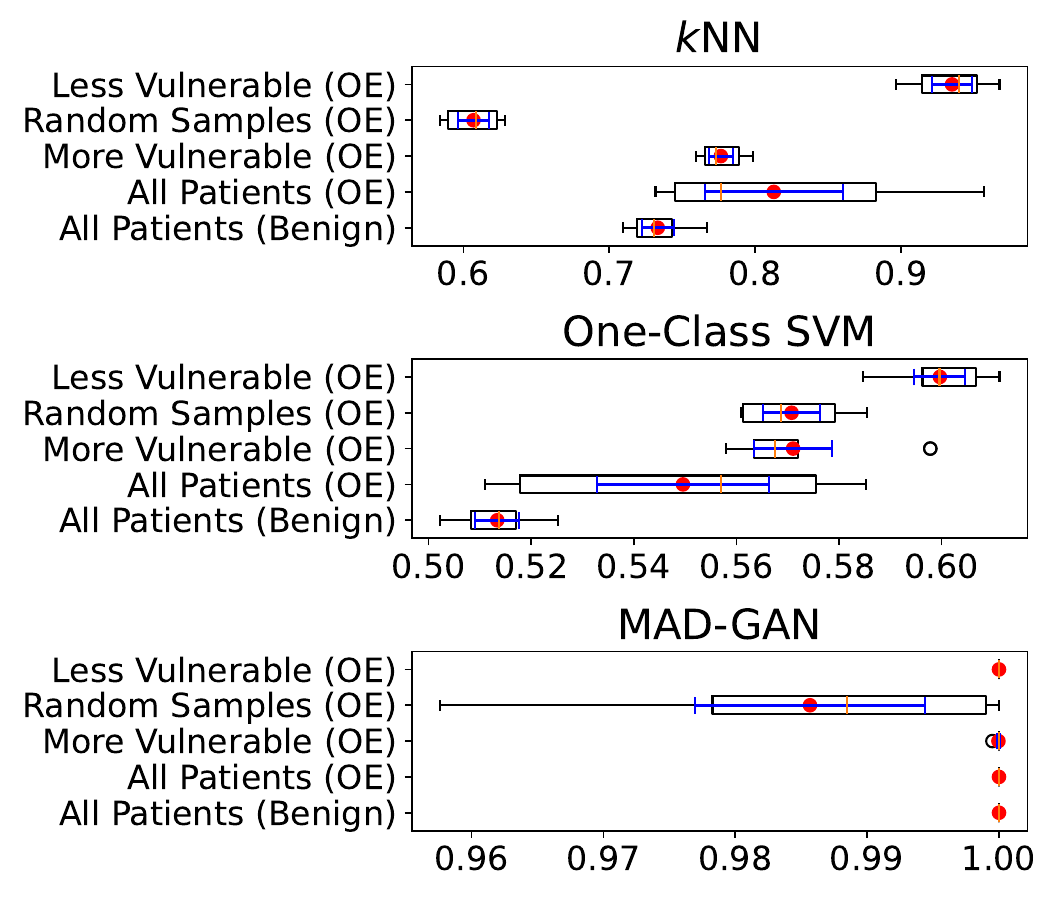}
			\end{minipage}}
			\hfill
			% Second subfigure
			\subfloat[MIMIC\label{Figure: MIMIC_Recall}]{\begin{minipage}{0.26\textwidth}
				\centering
				\includegraphics[width=\linewidth, trim=6.7cm 0 0 0, clip]{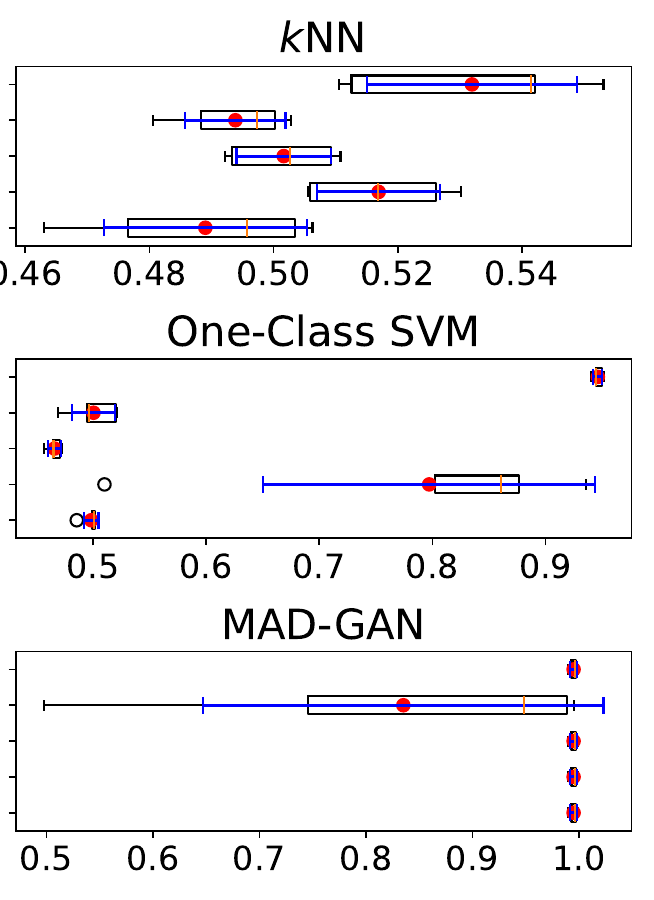}
			\end{minipage}}
			\hfill
			% Third subfigure
			\subfloat[PhysioNet CinC\label{Figure: Sepsis_Recall}]{\begin{minipage}{0.26\textwidth}
				\centering
				\includegraphics[width=\linewidth, trim=6.7cm 0 0 0, clip]{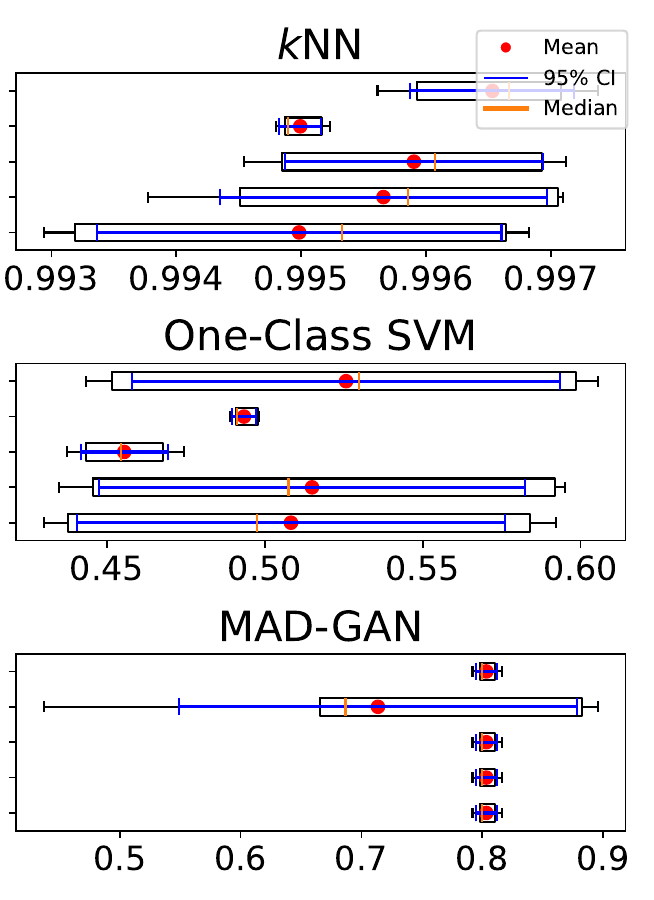}
			\end{minipage}}
		\end{minipage}
	}
	\caption{Recall of \textit{k}NN, One-Class SVM, and MAD-GAN. Less vulnerable training and OE (\systemname) improves recall by 27.8\% (\textit{k}NN-OhioT1DM), 89.8\% (One-Class SVM-MIMIC), and 3.13\% (One-Class SVM-PhysioNet CinC) vs. All Patients (Benign).}
	\label{Figure: Recall}
\end{figure*}

\textbf{Recall.} Figure \ref{Figure: Recall} shows that training on less vulnerable patients yields the highest recall across all datasets when exposed to the URET attack. 
% For each dataset, we use a paired t-test to compare \systemname with All Patients (Benign). Each patient contributes one matched pair of aggregate recall values (one under each training strategy) and the test is applied to the within-patient differences.
The largest relative recall gains are 27.8\% for \textit{k}NN on OhioT1DM, 
% (t-test = 17.4, p-value = $2.32 \times 10^{-9}$), 
89.8\% for One-Class SVM on MIMIC,
%  (t-test = 158.8, p-value = $9.44 \times 10^{-9}$), 
and 3.13\% for One-Class SVM on PhysioNet CinC.
% (t-test = 4.69, p-value = $9.38 \times 10^{-3}$). 
%All the improvements are statistically significant given that the p-values are less than 0.05. 
Some saturated cells are not informative for distinguishing training strategies. In particular, \textit{k}NN recall on PhysioNet CinC remains between 0.993 and 0.997 across the plotted strategies. 

Training on less vulnerable patients also outperforms training on the more vulnerable and random subsets. Thus, the gains are not merely a byproduct of reducing the training-set size; they are specific to selecting patients whose profiles are less sensitive to adversarial manipulation. Furthermore, less vulnerable training outperforms ``All Patients (OE)'', indicating that patient selection contributes beyond outlier exposure alone. For MAD-GAN, which has the highest all-patient benign-data recall on OhioT1DM and MIMIC, less vulnerable training matches full-dataset recall while substantially reducing the training-set size and training time.

Under FGSM, the recall gains follow similar trends, with average increases across all datasets of 9.90\% for \textit{k}NN, 
% (t-test = 5.87, p-value = $1.07 \times 10^{-4}$), 
1.88\% for One-Class SVM,
%  (t-test = 0.928, p-value = $0.373$), 
with no change for MAD-GAN, relative to ``All Patients (Benign)''. 
% While \textit{k}NN shows significant improvement, One-Class SVM shows a positive but not statistically significant change.

% \insightbox{RQ2 Insight \#1}{\systemname improves the recall of the evaluated ADs by 16.2\% and 3.93\% on average under the URET and FGSM attacks, respectively.}

\insightbox{RQ2 Insight \#1}{On average, \systemname improves the recall by 16.2\% across URET dataset--AD combinations and 3.93\% across FGSM dataset--AD combinations.}

The gains across ADs align with dataset characteristics: \systemname benefits \textit{k}NN most on the dense, low-to-moderate-dimensional OhioT1DM and One-Class SVM most on the sparser, higher-dimensional MIMIC and PhysioNet CinC. This pattern is consistent with \textit{k}NN's reliance on dense, stable neighborhoods and with the ability of margin-based detectors to operate on sparse, heterogeneous distributions.

Across the three datasets, peak recall gains follow the same ordering as normalized inter-cluster distance.
% : MIMIC has the largest distance and gain, PhysioNet CinC has the smallest, and OhioT1DM lies between them.
% Peak improvements also track cluster separation: 
Larger
% normalized inter-cluster 
distances produce purer, less vulnerable clusters and larger recall gains. MIMIC has the largest inter-cluster distance (37.84) and the largest gain (89.8\%), whereas PhysioNet CinC has the smallest distance (26.10) and the smallest gain (3.13\%). OhioT1DM falls
% neatly 
in between with a distance of 32.46 and a gain of 27.8\%. Because clustering is applied to patient risk profiles rather than raw physiological measurements, this distance captures separation in adversarial response rather than physical or clinical distance alone.

%\karthik{Can we actually relate cluster distances to physical quantities? Like what does it mean for two clusters to be far apart in the Ohio dataset? That they have different values of blood sugar?}

% \karthik{Isn't this also a function of the clustering algorithm?}
\insightbox{RQ2 Insight \#2}{Across the three datasets, the larger the inter-cluster distance, the higher the recall improvement achieved by \systemname.}

%\karthik{I think the precision should be a separate RQ as well}. 

\begin{figure}[t]
	\centering
	\includegraphics[width=\columnwidth]{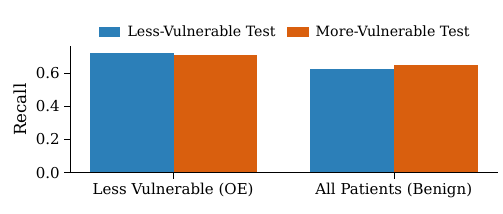}
	\caption{Recall on less-vulnerable and more-vulnerable test patients, averaged across datasets, attack algorithms, and ADs, for \systemname (Less Vulnerable (OE)) versus the All Patients (Benign) baseline. On average, \systemname achieves similar recall on both clusters and improves their recall over the baseline.}
	\label{Figure: PerCluster_Recall}
\end{figure}

\textbf{Fairness across patient clusters.} A natural concern is that excluding the more vulnerable patients from training shifts the benign distribution and leaves those patients under-protected at inference time, since their physiological patterns are no longer represented in the training set. To test this, we partition the test set by cluster and report recall separately on less-vulnerable and more-vulnerable test patients, averaged across the three datasets, four attack algorithms, and five ADs. Figure \ref{Figure: PerCluster_Recall} shows that \systemname (Less Vulnerable (OE)) sustains comparable average recall on both groups (0.72 and 0.71), and raises average recall on the more-vulnerable test patients relative to the All Patients (Benign) baseline (0.65 $\rightarrow$ 0.71). This indicates that training on the less vulnerable cluster does not degrade detection of adversarial samples targeting more vulnerable patients on average; rather, the more regular benign boundary learned by \systemname generalizes across both clusters.

\insightbox{RQ2 Insight \#3}{On average across datasets, attacks, and ADs, selectively training on less vulnerable patients does not under-protect more vulnerable ones. \systemname improves their detection recall over the baseline (0.65 $\rightarrow$ 0.71).}

\begin{figure*}[t]
	\centering
	\scalebox{0.9}{% scale the entire figure by 90%
		\begin{minipage}{\textwidth}
			\centering
			% First subfigure
			\subfloat[OhioT1DM\label{Figure: OhioT1DM_Precision}]{\begin{minipage}{0.42\textwidth}
				\centering
				\includegraphics[width=\linewidth]{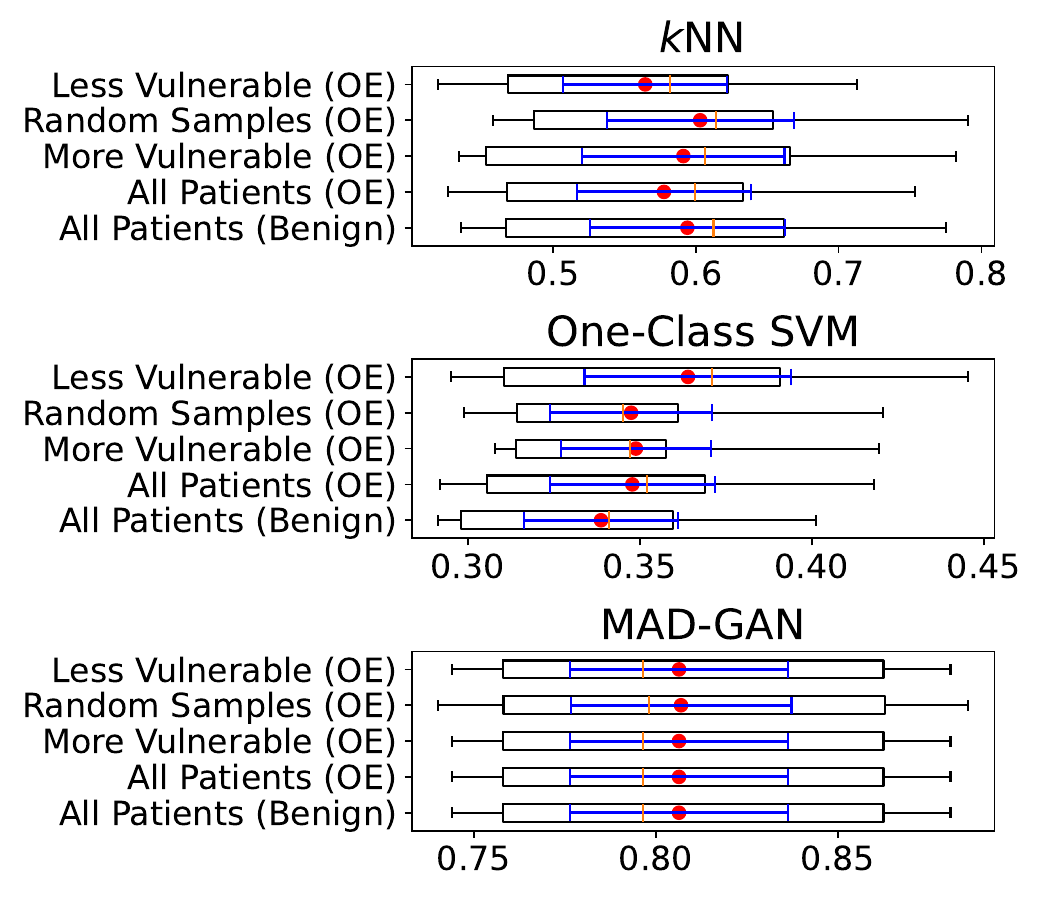}
			\end{minipage}}
			\hfill
			% Second subfigure
			\subfloat[MIMIC\label{Figure: MIMIC_Precision}]{\begin{minipage}{0.26\textwidth}
				\centering
				\includegraphics[width=\linewidth, trim=6.7cm 0 0 0, clip]{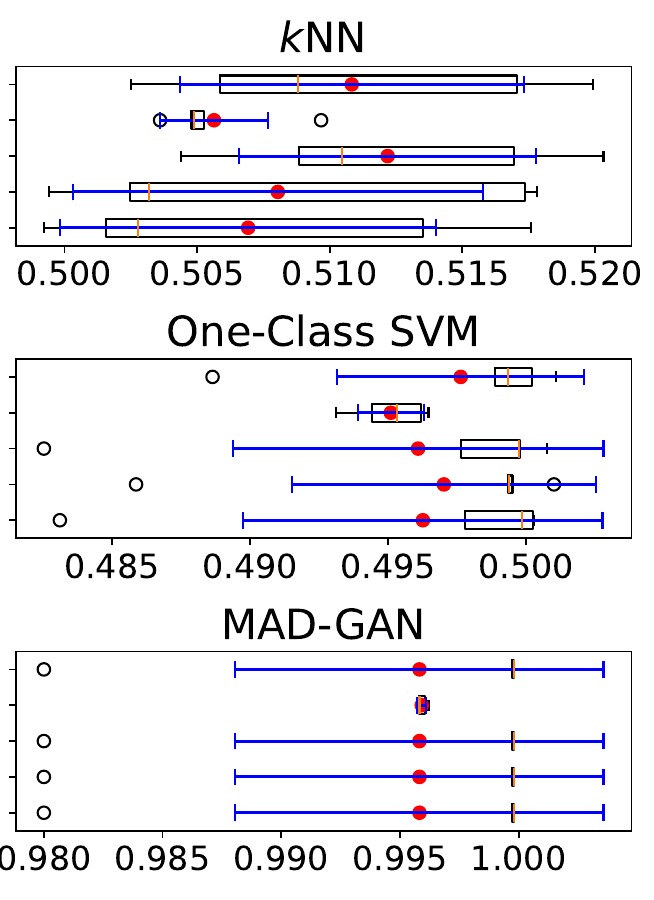}
			\end{minipage}}
			\hfill
			% Third subfigure
			\subfloat[PhysioNet CinC\label{Figure: Sepsis_Precision}]{\begin{minipage}{0.26\textwidth}
				\centering
				\includegraphics[width=\linewidth, trim=6.7cm 0 0 0, clip]{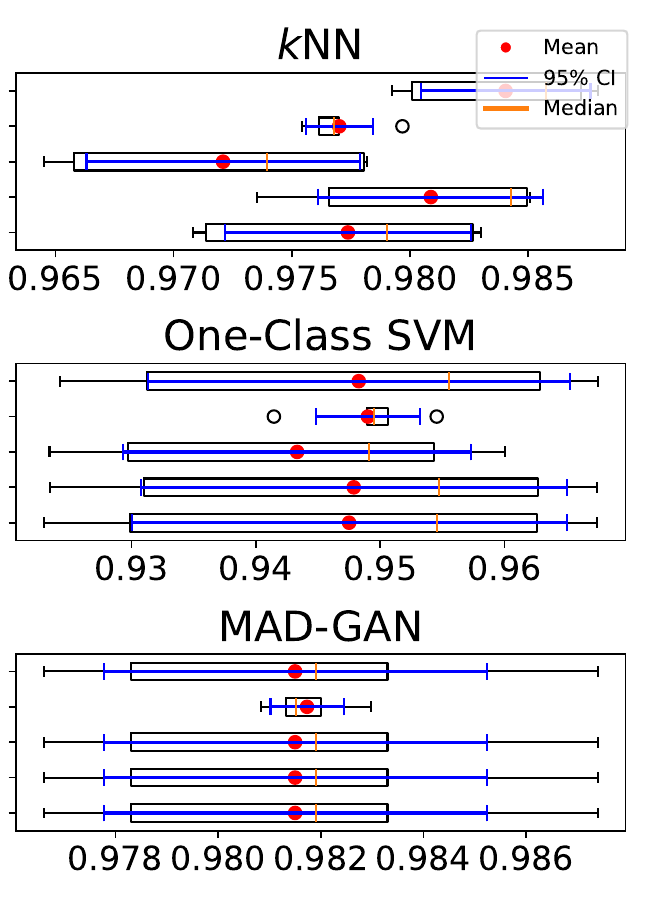}        
			\end{minipage}}
		\end{minipage}
	}
	\caption{Precision of \textit{k}NN, One-Class SVM, and MAD-GAN. Less vulnerable training and OE (\systemname) improves precision by 7.46\% (One-Class SVM-OhioT1DM), 0.77\% (\textit{k}NN-MIMIC), and 0.90\% (\textit{k}NN-PhysioNet CinC) vs. All Patients (Benign). Under URET, precision is preserved in all but one experiment.}
	\label{Figure: Precision}
\end{figure*}

\paragraph{\textbf{Answer to RQ3}}
\textbf{Precision.} Figure \ref{Figure: Precision} shows that precision drops in only one of the nine URET experiments: precision for \textit{k}NN on OhioT1DM decreases by 4.97\%
% (t-test = -4.95, p-value = $4.37 \times 10^{-4}$)
because excluding more vulnerable patients tightens local neighborhoods and slightly increases FPs for benign samples from excluded profiles. This effect is most visible for local, instance-based detectors such as \textit{k}NN on comparatively homogeneous data, where neighborhood composition changes more sharply after selective filtering. 
Two observations support this: (i) it is specific to \textit{k}NN since One-Class SVM and MAD-GAN show no precision drop on OhioT1DM, which is consistent with kNN's unique reliance on local neighborhood composition, a reliance not shared by the other defenses, and (ii) \textit{k}NN precision does not decrease on MIMIC or PhysioNet CinC, which have sparser and higher-dimensional data where removing a patient cluster changes neighborhood density less sharply than on the comparatively homogeneous OhioT1DM.

With the remaining defenses, selective training with outlier exposure maintains or improves precision. The largest precision gains are 7.46\% for One-Class SVM on OhioT1DM,
%  (t-test = 6.27, p-value = $6.08 \times 10^{-5}$)
0.77\% for \textit{k}NN on MIMIC, 
% (t-test = 2.56, p-value = $6.29 \times 10^{-2}$; $\alpha=0.05$)
and 0.90\% for \textit{k}NN on PhysioNet CinC. 
% (t-test = 7.66, p-value = $1.56 \times 10^{-3}$). 
% On MIMIC, the \textit{k}NN point estimate increases by 0.77\%.
% , but the difference is not significant (t-test = 2.56, p-value = $6.29 \times 10^{-2}$; $\alpha=0.05$); we therefore interpret this cell as approximately maintaining precision rather than as a precision gain. 
Overall, these results show that \systemname improves recall while largely preserving the detectors' ability to avoid false alarms.

The FGSM run provides an additional white-box check yielding similar detector-level patterns. One-Class SVM and MAD-GAN preserve precision under \systemname: One-Class SVM improves from 0.5365 to 0.5392, and MAD-GAN remains unchanged at 0.9996. The only primary-detector exception is \textit{k}NN, whose precision decreases from 0.6005 to 0.3932, reflecting the same local-neighborhood sensitivity observed under URET rather than a broader loss of precision across detector families. Thus, among the primary ADs, the FGSM results reinforce that precision preservation is strongest for the boundary-based and temporal generative detectors, while the local detector exposes the expected precision--recall tradeoff.

% \karthik{We should make it clear in both figures that outlier-exposed, less vulnerable training is what \systemname does.}

% The combination of selective filtering and controlled outlier exposure helps explain why precision usually remains stable or improves despite higher recall.
% Selecting the lower-impact patient cluster reduces cross-patient heterogeneity in the training set, while injecting adversarial samples near the benign boundary helps the AD learn a more precise separation between benign and adversarial behavior.
% In the single case where precision drops, namely \textit{k}NN on OhioT1DM, the local and instance-based nature of the AD makes it more sensitive to excluding benign neighborhoods associated with the removed patient profiles.
% \karthik{What evidence do we have for this?}
% Two observations support this: (i) it is specific to \textit{k}NN since One-Class SVM and MAD-GAN show no precision drop on OhioT1DM, which is consistent with kNN's unique reliance on local neighborhood composition unlike the other defenses, and (ii) \textit{k}NN does not drop on MIMIC or PhysioNet CinC, which have sparser and higher-dimensional data where removing a patient cluster changes neighborhood density less sharply than on the comparatively homogeneous OhioT1DM.
%This failure mode is narrow rather than systemic, since it does not recur across the other datasets or detectors.
%\karthik{Ok, but what happened in that one case?}
%\karthik{Can we explain why we saw a slight increase in the precision? Also, why did the precision decrease in the one case}

\insightbox{RQ3 Insight}{\systemname preserves precision for boundary-based and temporal generative detectors; the only observed precision loss is confined to the local, instance-based \textit{k}NN detector.}

\paragraph{\textbf{Answer to RQ4}}
By eliminating more vulnerable patients and using less vulnerable ones only for selective training, \systemname reduces the number of patients in the training set by 75.0\% for OhioT1DM, 93.2\% for MIMIC, and 72.0\% for PhysioNet CinC, for an average reduction of 80.1\%. This reduction is especially important for the larger, more heterogeneous MIMIC and PhysioNet CinC datasets, where indiscriminate training would incur a substantial computational cost. 
	
Table \ref{Table: Training_Time} shows the corresponding reduction in AD training time for One-Class SVM and MAD-GAN. Since \textit{k}NN is instance-based and defers computation to inference time, we omit it. The average per-cycle training-time reduction is 88.3\%. This percentage describes detector training only and does not include \systemname's one-time overhead, which comprises attack simulation, risk profiling, and clustering.
%\karthik{Where do we mention the training set size reduction? We should remove it from the insight and the RQ if not.}

% Please add the following required packages to your document preamble:
% \usepackage{multirow}
% \usepackage{graphicx}
\begin{table}[t]
	\centering
	\caption{AD training times (s) under All Patients (Benign) vs. Less Vulnerable OE (\systemname) training.}
\label{Table: Training_Time}
\resizebox{\columnwidth}{!}{%
	\begin{tabular}{|l|lll|lll|}
		\hline
		\multicolumn{1}{|c|}{\multirow{2}{*}{\textbf{Dataset}}} &
		\multicolumn{3}{c|}{\textbf{One-Class SVM}} &
		\multicolumn{3}{c|}{\textbf{MAD-GAN}} \\ \cline{2-7} 
		\multicolumn{1}{|c|}{} &
		\multicolumn{1}{l|}{\textbf{\begin{tabular}[c]{@{}l@{}}All \\ Patients \\ (Benign)\end{tabular}}} &
		\multicolumn{1}{l|}{\textbf{\begin{tabular}[c]{@{}l@{}}Less \\ Vulnerable \\ (OE)\end{tabular}}} &
		\textbf{\begin{tabular}[c]{@{}l@{}}Percentage \\ Decrease \end{tabular}} &
		\multicolumn{1}{l|}{\textbf{\begin{tabular}[c]{@{}l@{}}All \\ Patients \\ (Benign)\end{tabular}}} &
		\multicolumn{1}{l|}{\textbf{\begin{tabular}[c]{@{}l@{}}Less \\ Vulnerable \\ (OE)\end{tabular}}} &
		\textbf{\begin{tabular}[c]{@{}l@{}}Percentage \\ Decrease \end{tabular}} \\ \hline
		OhioT1DM &
		\multicolumn{1}{l|}{472.31} &
		\multicolumn{1}{l|}{27.98} &
		94.08 &
		\multicolumn{1}{l|}{1732.12} &
		\multicolumn{1}{l|}{421.56} &
		75.66 \\ \hline
		MIMIC &
		\multicolumn{1}{l|}{19551.84} &
		\multicolumn{1}{l|}{45.74} &
		99.77 &
		\multicolumn{1}{l|}{10384.41} &
		\multicolumn{1}{l|}{272.69} &
		97.37 \\ \hline
		PhysioNet CinC &
		\multicolumn{1}{l|}{54173.10} &
		\multicolumn{1}{l|}{4224.19} &
		92.20 &
		\multicolumn{1}{l|}{1705.10} &
		\multicolumn{1}{l|}{500.34} &
		70.66 \\ \hline
	\end{tabular}%
}
\end{table}

\textbf{\systemname Overhead.} We evaluate \systemname's one-time overhead by reporting the execution time per patient of each step in Table \ref{Table: Execution_Time}. Attack simulation on the OhioT1DM model takes approximately $625\times$ and $1732\times$ as long as attack simulation on the MIMIC and PhysioNet CinC models, respectively. This difference arises because the deep residual time-series forecasting architecture produces multi-step forecasts, whereas the other models produce simpler single-step forecasts.
% \karthik{Provide the numbers here} 
%	\karthik{Can you phrase this less negatively? How about saying it takes much longer instead?}
Moreover, clustering time increases with the number of patients in the dataset.

	% Please add the following required packages to your document preamble:
	% \usepackage{graphicx}
	\begin{table}[t]
		\centering
		\caption{Per-patient execution time (s) for \systemname's attack simulation, risk profiling, and clustering steps across datasets.}
		\label{Table: Execution_Time}
		\resizebox{\columnwidth}{!}{%
			\begin{tabular}{|l|l|l|l|}
				\hline
				\textbf{Dataset} & \textbf{Attack Simulation} & \textbf{Risk Profiling} & \textbf{Clustering} \\ \hline
				OhioT1DM       & 1525.64 & 0.000419 & 0.0491 \\ \hline
				MIMIC          & 2.44 & 0.000963 & 0.528 \\ \hline
				PhysioNet CinC & 0.881 & 0.00011 & 2.73 \\ \hline
			\end{tabular}%
		}
	\end{table}

% Because \systemname is applied before deployment, it incurs a one-time overhead and does not affect DNN inference time. The added cost is therefore operationally manageable even when attack simulation is expensive, particularly because clustering and profiling   costs are negligible in most cases relative to attack generation.
\insightbox{RQ4 Insight}{On average, \systemname reduces the number of training patients by 80.1\% and AD training time by 88.3\% per training cycle, excluding \systemname's one-time attack-simulation, risk-profiling, and clustering overhead.}

\begin{figure}[t]
	\centering
	\subfloat[All Patients (OE) baseline\label{Figure: CrossAttack_Baseline}]{\begin{minipage}{0.49\columnwidth}
		\centering
		\includegraphics[width=\linewidth]{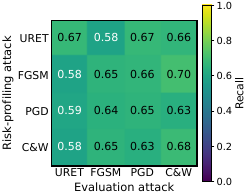}
	\end{minipage}}
	\hfill
	\subfloat[Less Vulnerable (OE) (\systemname)\label{Figure: CrossAttack_System}]{\begin{minipage}{0.49\columnwidth}
		\centering
		\includegraphics[width=\linewidth]{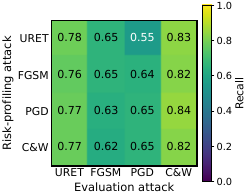}
	\end{minipage}}
	\caption{Aggregate cross-attack recall across the three datasets and five ADs. 
	% Rows indicate the profiling attack and columns the evaluation attack; off-diagonal cells use different attacks. 
	\systemname's mean off-diagonal recall remains close to its mean diagonal recall indicating that clusters generalize beyond the profiling attack on average.}
	\label{Figure: CrossAttack}
\end{figure}

\paragraph{\textbf{Answer to RQ5}}
Because \systemname uses an evasion attack both to profile patient vulnerability and to evaluate recall, a reasonable concern is that the observed gains are circular, i.e., specific to the attack used for clustering. To examine this concern, we perform a cross-attack analysis over four attacks spanning both threat models: URET (black-box), FGSM, PGD, and C\&W (white-box). We profile patients and form clusters using each attack in turn, then evaluate recall against each attack. Figure \ref{Figure: CrossAttack} reports recall averaged across the three datasets and five ADs for the All Patients (OE) baseline (Figure \ref{Figure: CrossAttack_Baseline}) and for \systemname (Figure \ref{Figure: CrossAttack_System}).
Two observations emerge. \textit{First}, \systemname's mean off-diagonal recall ($\approx 0.71$) remains close to its mean diagonal recall ($\approx 0.73$), showing that, on average across datasets and ADs, clusters formed under one attack transfer to other attacks rather than overfitting to the profiling attack. \textit{Second}, \systemname exceeds the corresponding baseline in most attack pairs, including evaluations with the multi-step PGD and C\&W attacks. PGD is the weakest evaluation column, for which \systemname obtains an average recall of 0.62 versus 0.65 for the baseline, a drop of 4.62\% relative to the baseline.

\insightbox{RQ5 Insight}{On average across datasets and ADs, \systemname's recall gains generalize across attacks as clusters formed with one attack transfer to others.}

\section{Threats to Validity}
\label{Section: Threats_to_Validity}

\textbf{Threats to Internal Validity.} There are three threats to internal validity. 

\textit{First}, \systemname 
	%has four main limitations. \textit{First}, it 
assumes that training and testing data are drawn from the same distribution, thereby overlooking distribution shifts.
	%~\cite{zhang2022towards}. 
%	Concept drift refers to the change in the statistical properties of patient data (e.g., the relationships between vital signs and outcomes) over time, which poses a challenge to the generalizability of \systemname to different data distributions and its adaptability to varying environments.
Distribution shift is a general challenge for ADs, and 
\systemname does not introduce additional shift sensitivity. 
We address short-term shifts through chronological splits ($\approx 80\%$ early samples for risk profiling, clustering, and AD training and $\approx 20\%$ later samples for evaluation), which expose the models to natural short-term physiological changes without temporal leakage. The available datasets do not span sufficiently long periods to support a meaningful evaluation of longer-term shifts caused by recovery or treatment changes.
%	In deployment, drift detection techniques such as monitoring shifts in risk-score distributions can trigger periodic risk profiling.
	% \gargi{Is this a proven fact that it's going to fail? Or a speculation? We can say that generalization might be a challenge, and in the absence of a suitable dataset, we haven't been able to test it. Also, what does concept drift mean in the context of healthcare?}
	%For instance, a model trained on senior patients may perform poorly on younger ones. This is difficult to verify because healthcare datasets are typically anonymized to protect privacy.
	% For example, a risk profiler trained on the data of senior patients may not be as accurate with young patients. We found this difficult to verify, since healthcare datasets are often anonymized to preserve patient privacy.
	% \gargi{We can say that we might need to split patients according to their demographic and compute their risk profiles separately. Bottomline: Instead of saying upfront it's going to fail, it's better to mention associated challenges and potential solutions, and why we haven't been able to implement them just yet. That would eventually lead to future work.}.
	
\textit{Second}, \systemname depends on the instantaneous impact formula used to build patient risk profiles; poor impact quantification may produce impure clusters and weaken the quality of the less vulnerable subset used for training. For this reason, we recommend consulting a domain expert when defining the impact proxy. %This design choice reflects the broader fact that domain knowledge remains central to risk-aware ML in healthcare.
In this work, however, the nonnegative regression-derived weights are used only as an empirical proxy for adversarial influence. Thus, \systemname is a systems-security method for improving attack detection and does not require clinically validated weights to function. 
	% , and the sensitivity analysis in Section~\ref{Section: Evaluation} shows that the resulting cluster assignments remain stable under moderate weight variation.
	
\textit{Third}, as discussed in §\ref{Section: Motivational_Case_Study}, \systemname assumes that lower attack-induced impact is associated with traces that are more suitable for selective AD training. 
% The motivational case study establishes cross-patient differences in statistical outlier prevalence but does not establish a cause of attack vulnerability. In domains where attack-induced impact does not align with benign-distribution heterogeneity, selective training may introduce bias. 
If attack-induced impact does not align with benign-distribution heterogeneity in another domain, selective training may introduce bias.
Evaluating this possibility is an avenue for future work.
	
\textbf{Threats to External Validity.} There are four threats to external validity. 
	
\textit{First}, \systemname uses offline training to build a static risk profiler, which may require periodic updates as patient profiles or attack environments change.
	% dataset changes. For example, patients move from high-risk to low-risk categories 
Such updates require rerunning attack simulation, risk profiling, and clustering, whose costs are strongly dataset-dependent, and are dominated by attack simulation. Profile updates should therefore be scheduled as explicit maintenance operations  and can be integrated into routine model maintenance cycles commonly used in healthcare ML systems.
	% Because updates operate on aggregated patient-level risk statistics rather than requiring full AD retraining, this update process is lightweight and can be integrated into routine model maintenance cycles commonly used in healthcare ML systems. 
%	\textit{Second}, we used a single attack algorithm to test the efficacy of \systemname. Although we chose URET due to its effectiveness and good representation of evasion attacks, more algorithms are needed for a more thorough evaluation. 

\textit{Second}, although our main recall and precision results use the black-box URET framework and the white-box FGSM attack, the cross-attack analysis (RQ5, Figure \ref{Figure: CrossAttack}) extends the evaluation to PGD and C\&W and shows that the benefit of selective training is not tied to the attack used for risk profiling. 
However, the evaluated attacks are AD-oblivious since none jointly optimizes the victim-model loss and the AD's anomaly score. 
An AD-aware adversary could potentially reduce detection recall by explicitly trading off attack success against detectability, particularly for differentiable detectors such as Deep SVDD, LSTM-AE, and MAD-GAN's discriminator. 
Evaluating such an adversary using a joint-objective attack remains valuable future work.

% Evaluating additional or adaptive attack families remains valuable future work.
%	 \karthik{Can we say that our algorithm is a pretty representative one?} 

\textit{Third}, we currently manipulate one vital sign per dataset to ensure the setup is realistic and the results are easy to interpret. %We treat this as a conservative starting point. 
However, \systemname is designed to support multi-feature attacks.

\textit{Fourth}, the OhioT1DM dataset comprises only 12 patients, which limits the generalizability of the clustering results for that dataset. MIMIC and PhysioNet CinC, with thousands of patients, provide broader coverage.
%	\karthik{We should put a more positive spin on this - perhaps say we considered this as a starting point but it can be generalized to other attacker models.}

%\karthik{How are these different from the limitations of \systemname in the next section? Combine these together.}

%\karthik{Structure this as a proper "Threats to Validity" section. Again, Abraham or Rui can help with this.}

\section{Conclusion}
\label{Section: Conclusion}

We present \systemname, a novel risk-aware selective training strategy with outlier exposure for anomaly detectors (ADs) against evasion attacks in healthcare.
% {\color{red}\systemname consists of two main stages: (1) risk profiling, which groups patients into different clusters depending on their level of vulnerability to evasion attacks, and (2) AD enhancement, which selectively trains ADs on benign and adversarial data from less vulnerable patients to improve AD recall without compromising their precision. While training on the less vulnerable patients aims to improve recall by reducing the noise arising from the most vulnerable patient data due to their erratic physiological behaviour, outlier exposure aims to reduce the impact on precision by carefully injecting adversarial samples into the less vulnerable patients' data to reduce the false positive rate.}
% \gargi{This looks more like a summary than conclusion. Also, we shouldn't at this point use the term `aims to', as we have already shown that the aim has been achieved. Instead, this part can be rephrased in the lines of: \systemname has two main features -- it trains the AD on the least vulnerable patients with a controlled amount of exposed outliers.} 
Across three healthcare datasets, \systemname improves recall by 16.2\% and 3.93\% on average under the black-box and white-box settings, respectively, relative to indiscriminate training, while largely preserving precision. On average, it reduces training time by 88.3\% per training cycle, excluding \systemname's one-time attack-simulation, risk-profiling, and clustering overhead. These results demonstrate the potential of risk-aware selective training to improve adversarial-sample detection while reducing training costs.

%\karthik{Again, explain what's the recall you're improving, and also present the average value, not best case.}

% \gargi{End with a couple of sentences on the broader impact this would have in the healthcare domain.}

In future work, we plan to build an online risk profiler that accounts for changing attack environments, algorithms, and data distributions. We also plan to test \systemname on personalized DNNs and evaluate its generalizability to other safety-critical domains, such as autonomous vehicles.
%\karthik{Move to the  conclusions section.}

% \section*{Acknowledgments}

\bibliographystyle{IEEEtran}
\bibliography{cite}

% \section{Biography Section}
% If you have an EPS/PDF photo (graphicx package needed), extra braces are
%  needed around the contents of the optional argument to biography to prevent
%  the LaTeX parser from getting confused when it sees the complicated
%  $\backslash${\tt{includegraphics}} command within an optional argument. (You can create
%  your own custom macro containing the $\backslash${\tt{includegraphics}} command to make things
%  simpler here.)
 
% \vspace{11pt}

% \bf{If you include a photo:}\vspace{-33pt}
% \begin{IEEEbiography}[{\includegraphics[width=1in,height=1.25in,clip,keepaspectratio]{fig1}}]{Michael Shell}
% Use $\backslash${\tt{begin\{IEEEbiography\}}} and then for the 1st argument use $\backslash${\tt{includegraphics}} to declare and link the author photo.
% Use the author name as the 3rd argument followed by the biography text.
% \end{IEEEbiography}

% \vspace{11pt}

% \bf{If you will not include a photo:}\vspace{-33pt}
% \begin{IEEEbiographynophoto}{John Doe}
% Use $\backslash${\tt{begin\{IEEEbiographynophoto\}}} and the author name as the argument followed by the biography text.
% \end{IEEEbiographynophoto}

% \vfill

\end{document}